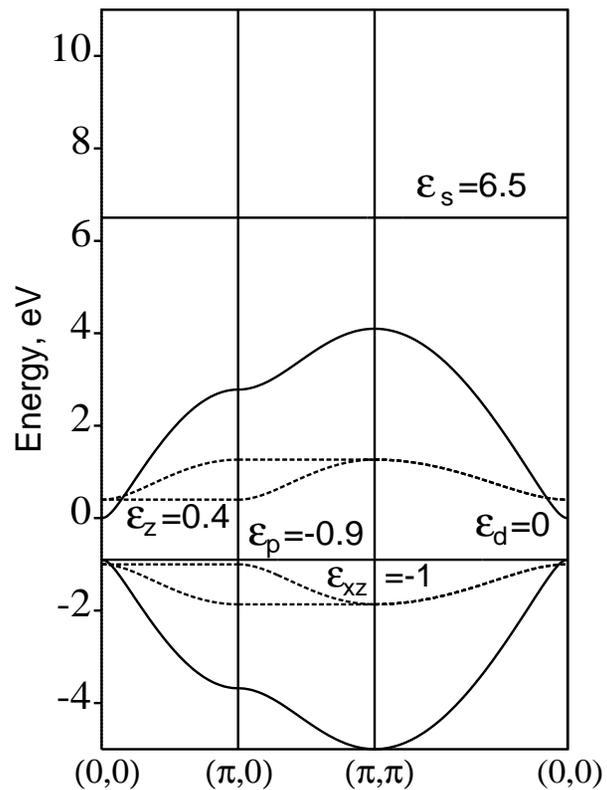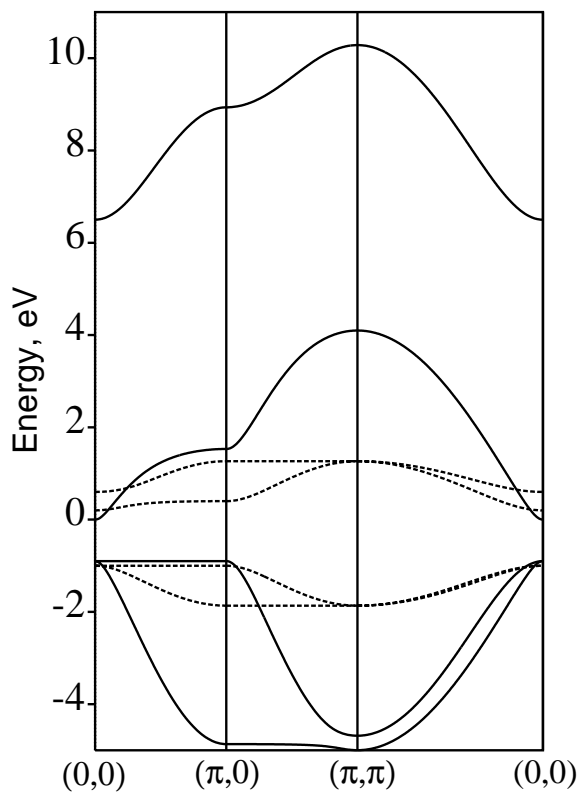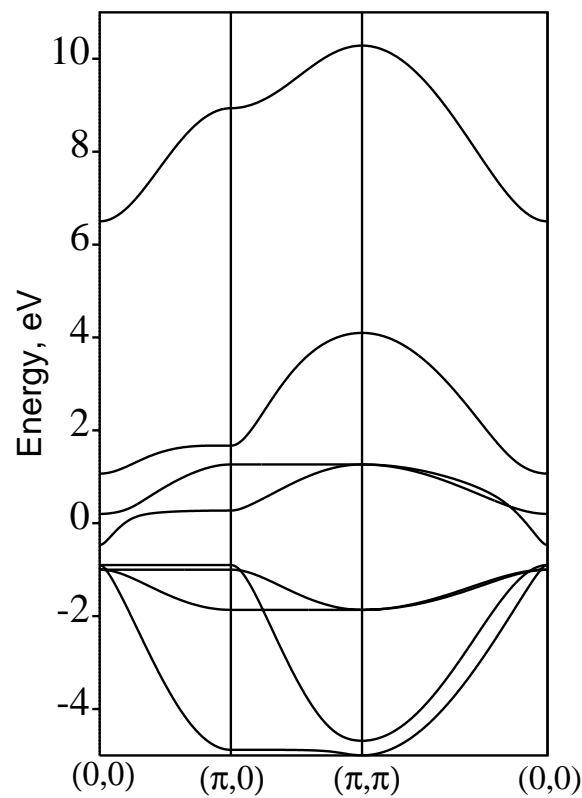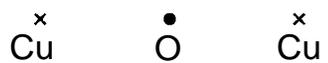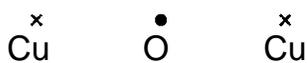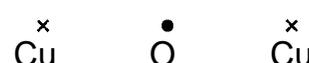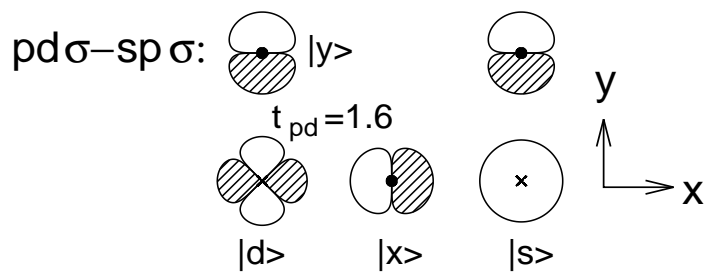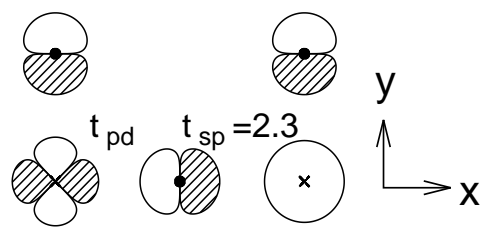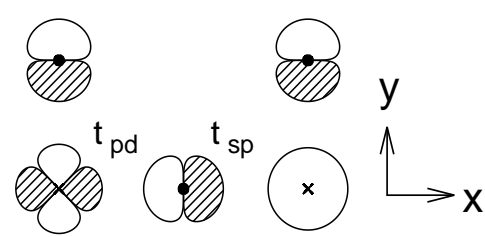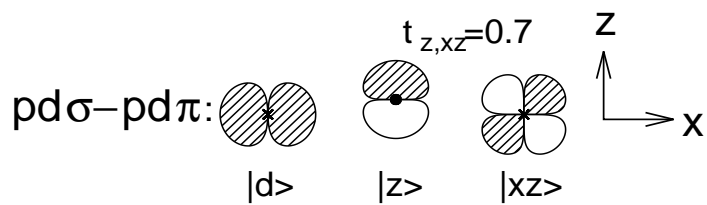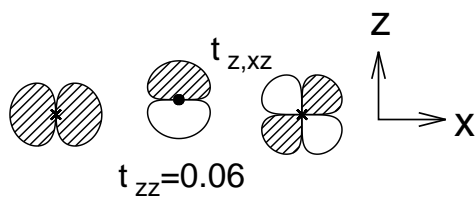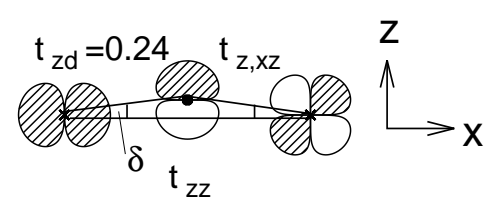

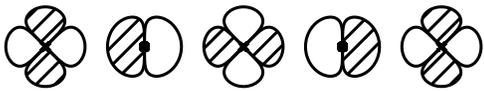
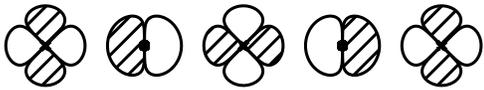

$X=(\pi,0)$ → $k$

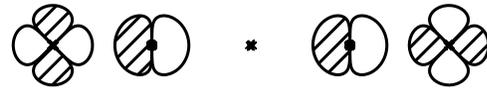
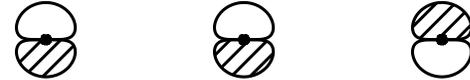
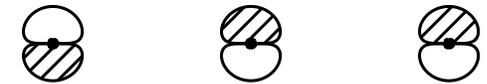
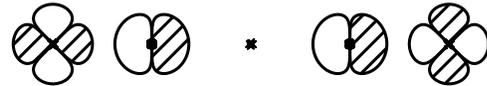

$\Gamma M/2 = (\pi/2,\pi/2)$ → $k$

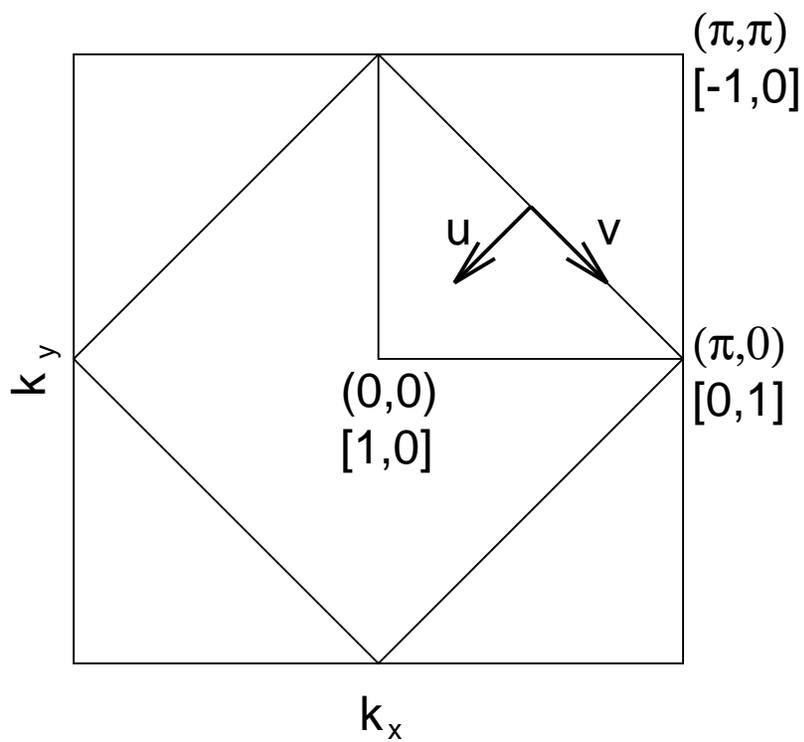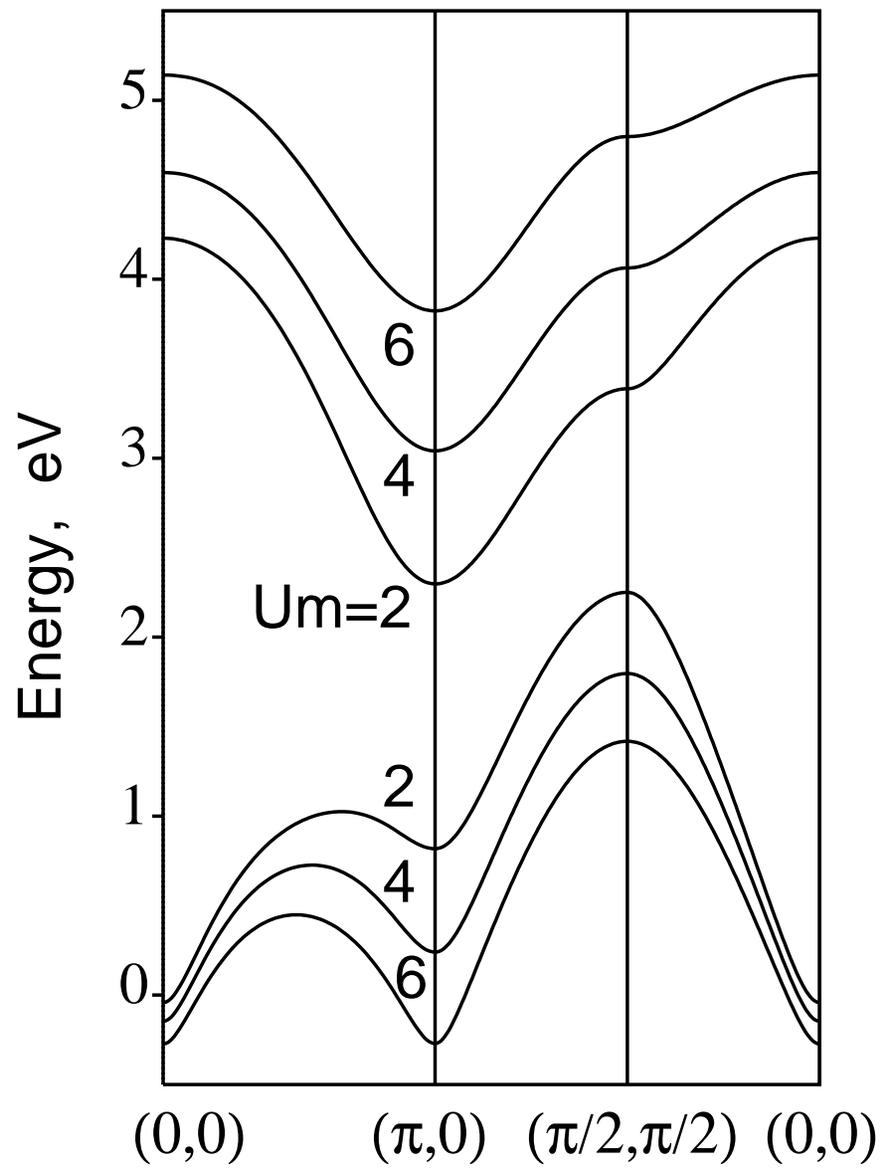

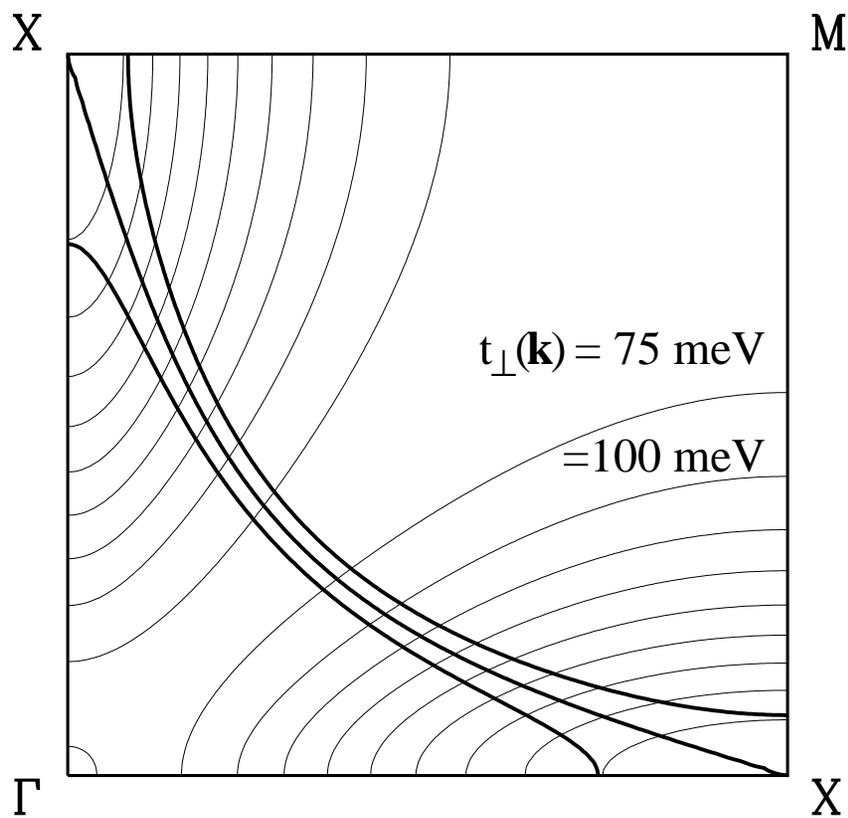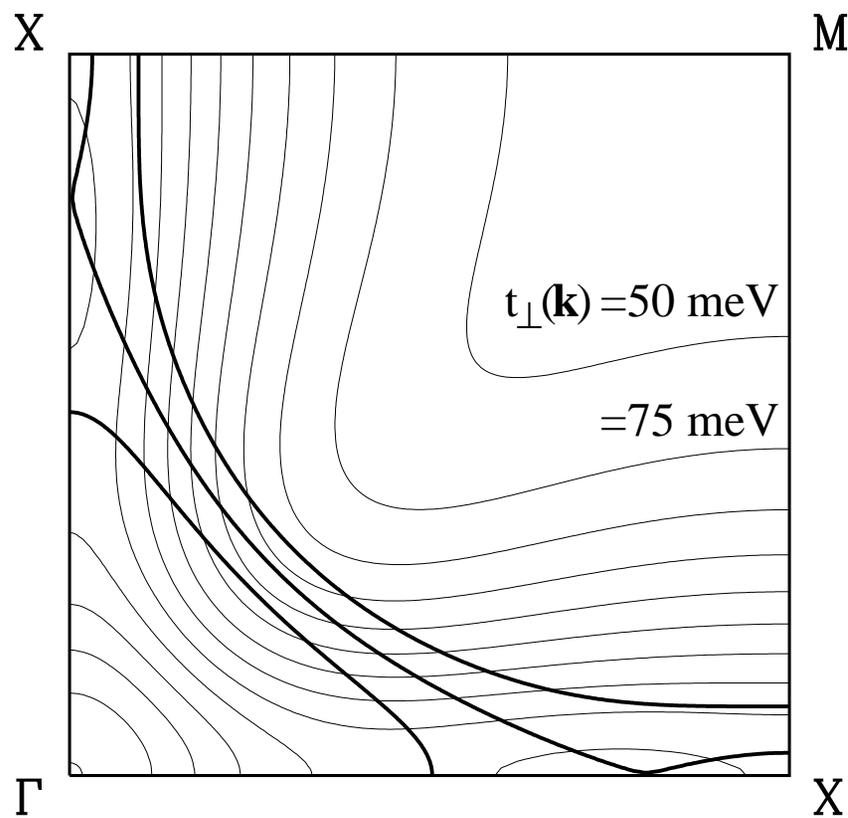

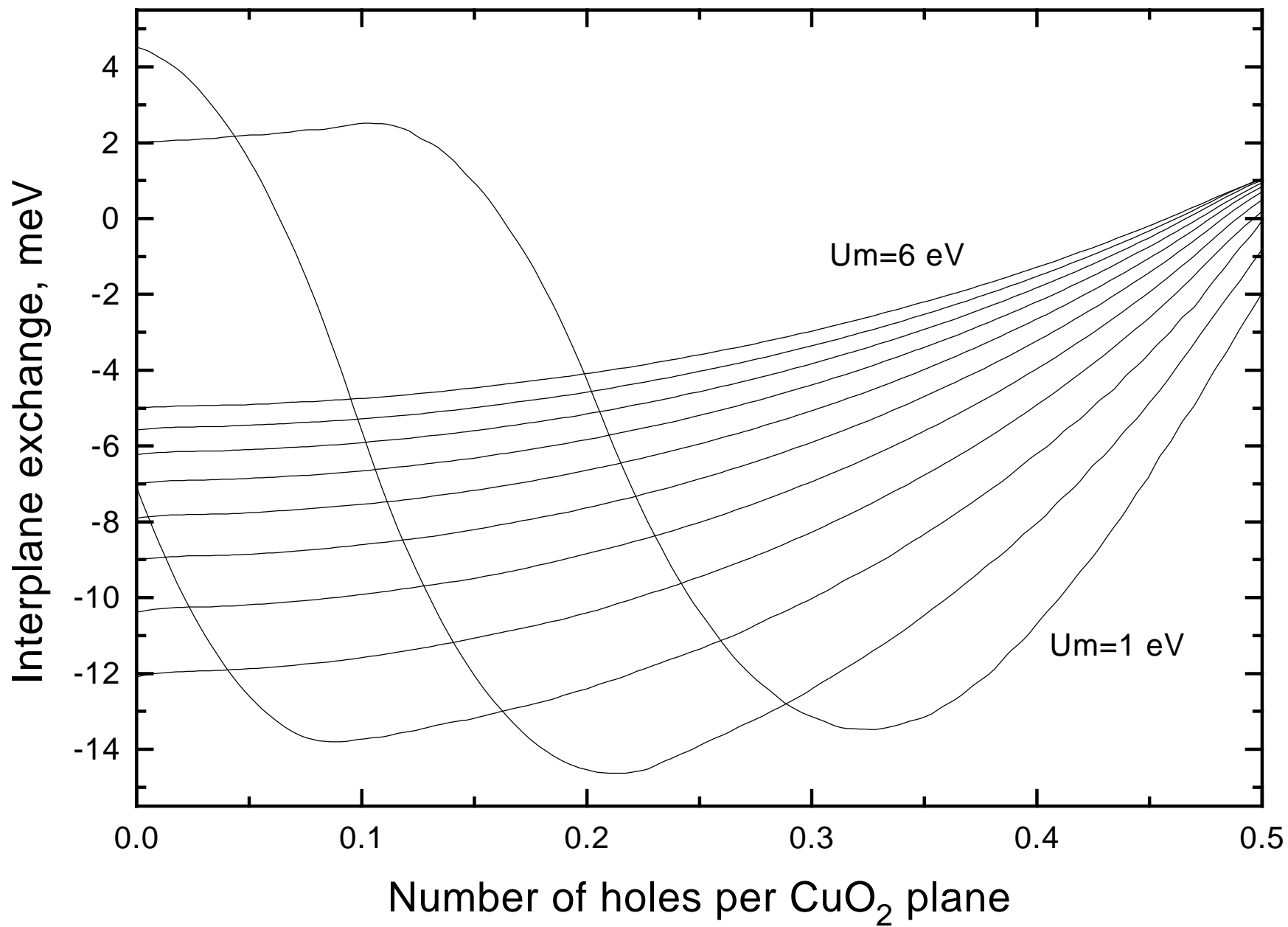

# LDA energy bands, low-energy Hamiltonians, $t'$, $t''$, $t_\perp(\mathbf{k})$, and $J_\perp$.

O. K. Andersen, A. I. Liechtenstein, O. Jepsen, and F. Paulsen.
Max-Planck Institut für Festkörperforschung, D-70569 Stuttgart, F.R.G.

## Abstract

We describe the LDA bandstructure of $YBa_2Cu_3O_7$ in the $\epsilon_F \pm 2$ eV range using orbital projections and compare with $YBa_2Cu_4O_8$. Then, the high-energy and chain-related degrees of freedom are integrated out and we arrive at two, nearest-neighbor, orthogonal, two-center, 8-band Hamiltonians, $H_8^+$ and $H_8^-$, for respectively the even and odd bands of the bi-layer. Of the 8 orbitals, $Cu_{x^2-y^2}$, $O2_x$, $O3_y$, and $Cu_s$ have $\sigma$ character and $Cu_{xz}$, $Cu_{yz}$, $O2_z$, and $O3_z$ have $\pi$ character. The roles of the $Cu_s$ orbital, which has some $Cu_{3z^2-1}$ character, and the four $\pi$ orbitals are as follows: $Cu_s$ provides 2nd- and 3rd-nearest-neighbor ($t'$ and $t''$) intra-plane hopping, as well as hopping between planes ($t_\perp$). The $\pi$-orbitals are responsible for bifurcation of the saddle-points for dimpled planes. The 4-$\sigma$-band Hamiltonian is generic for flat $CuO_2$ planes and we use it for analytical studies. The $\mathbf{k}_\parallel$-dependence is expressed as one on $u \equiv (\cos bk_y + \cos ak_x)/2$ and one on $v \equiv (\cos bk_y - \cos ak_x)/2$. The latter arises solely through the influence of $Cu_s$. The reduction of the $\sigma$-Hamiltonian to 3- and 1-band Hamiltonians is explicitly discussed and we point out that, in addition to the hoppings commonly included in many-body calculations, the 3-band Hamiltonian should include hopping between *all* 2nd-nearest-neighbor oxygens and that the 1-band Hamiltonian should include 3rd-nearest-neighbor hoppings. We calculate the single-particle hopping between the planes of a bi-layer and show that it is generically: $t_\perp(\mathbf{k}_\parallel) \approx 0.25$ eV$\cdot v^2 (1 - 2ut'/t)^{-2}$. The hopping through insulating spacers such as (BaO)Hg(BaO) is an order of magnitude smaller, but seems to have the same $\mathbf{k}_\parallel$-dependence. We show that the inclusion of $t'$ is crucial for understanding ARPES for the anti-ferromagnetic insulator $Sr_2CuO_2Cl_2$. Finally, we estimate the value of the inter-plane exchange constant $J_\perp$ for an un-doped bi-layer in mean-field theory using different single-particle Hamiltonians, the LDA for $YBa_2Cu_3O_6$, the eight- and four-band Hamiltonians, as well as an analytical calculation for the latter. We conclude that $J_\perp \sim -20$ meV.

# 1 Introduction.

For the HTSC materials practitioners of the LDA now basically agree about their results. The major uncertainties come from the compositional and structural data rather than from the computational techniques. For the various materials, the energy bands originating from the $CuO_2$-planes are fairly similar, merely the position of the Fermi level changes.

For stoichiometric HTSC's such as $YBa_2Cu_3O_7$ the LDA has been successful in reproducing or predicting normal-state structural and phononic properties [1, 2, 3], optical excitation spectra at high energies ($\hbar\omega > 1$ eV for metals, $\hbar\omega > 3$ eV for Mott-Hubbard insulators) [4], and $k_z$-averaged Fermi surfaces $\epsilon(\mathbf{k}_\parallel)=0$ for the metals [5, 3]. The most surprising LDA prediction for HTSC's with dimpled $CuO_2$ planes ($YBa_2Cu_3O_7$ [1, 3], $YBa_2Cu_3O_{6.5}$ [6]) has been that of bifurcated saddle-points close to the Fermi level. *Extended* saddle-points near $\epsilon_F$ were subsequently found in ARPES [7], but whether these are related to the bifurcated LDA saddle-points through "Fermion condensation" [8], or are caused by anti-ferromagnetic spin-fluctuations [9, 10], or are due to surface effects [11], remains to be seen. The LDA has finally provided reasonable estimates of the electron-electron (e.g. $U_{Cu\,d} \sim 7$ eV) [12, 13, 14] and electron-phonon interactions (e.g. $\lambda_{e-ph} \sim 1$ and softening of the Raman-active dimpling modes for $T < T_C$) [1, 15, 3, 16]. $\lambda_{e-ph} \sim 1$ might account for $T_c < 40$ K, but hardly for 100 K.

However, the LDA band structures for HTSC's are complicated, rarely interpreted, not accurate – or even relevant – below 50 meV, and are not delivered in a form useful as the single-particle term of a correlated model Hamiltonian which describes the low-energy excitations. As a result, most theorists neglect the LDA band structure, or at least its non-trivial details, and use the simplest possible Cu-Cu one-band, two-center, orthogonal, tight-binding model with hopping integrals $t$ and $t'$ between only nearest and next-nearest Cu neighbors (see Eq. (16)). A few brave theorists [10, 9] use the Emery three-band model.

We have therefore integrated out of the LDA band structure for $YBa_2Cu_3O_7$ the high-energy and chain-related degrees of freedom. In a first step we used the LMTO downfolding technique [17] to derive a two-dimensional, *nearest-neighbor*, two-center, orthogonal, tight-binding Hamiltonian with *eight* orbitals per $CuO_2$ plane [18]. Of these eight $CuO_2$ orbitals, four ($Cu_{x^2-y^2}$, $O2_x$, $O3_y$, and $Cu_s$) have $\sigma$ character and four ($Cu_{xz}$, $Cu_{yz}$, $O2_z$, and $O3_z$) have $\pi$ character with respect to the plane. Comparisons with full LDA bands calculated for other structurally well-characterized HTSC materials convinced us, that this eight-band Hamiltonian is generic for LDA plane bands.

The present paper starts with a qualitative description of the *full* LDA band structure for $YBa_2Cu_3O_7$ and $YBa_2Cu_4O_8$ in the $\epsilon_F \pm 2$ eV region and emphasizes the *chain*-related features. Then we proceed with a quantitative description of the *plane bands* by means of the eight-band Hamiltonian. The roles of the $Cu_s$ orbital, which possesses some $Cu_{3z^2-1}$ character, and the four $\pi$-orbitals are specifically explained: $Cu_s$ provides 2nd- and 3rd-nearest-neighbor ($t'$ and $t''$) intra-plane hopping, as well as hopping between planes ($t_\perp$). The $\pi$-orbitals are responsible for bifurcation of the saddle-points. We explicitly show how further high-energy degrees of freedom may be integrated out. The four $\sigma$-bands are studied analytically. Their $\mathbf{k}_\parallel$-dependence is most simply expressed as one on $u \equiv (\cos bk_y + \cos ak_x)/2$ and one on $v \equiv (\cos bk_y - \cos ak_x)/2$, where the latter arises solely through the influence of $Cu_s$. The reduction of the $\sigma$-band Hamiltonian to the commonly used 3- and 1-band Hamiltonians is explicitly discussed, and we point out that, to be consistent with the LDA bands, the former should include hopping between *all* 2nd-nearest-neighbor oxygens and the latter should include 3rd-nearest-neighbor hopping. As a first simple application, we calculate the single-particle hopping between the planes of a bilayer and show that it is generically $t_\perp(\mathbf{k}_\parallel) \sim 0.25$ eV$\cdot v^2 (1 - 2ut'/t)^{-2}$. The hopping through insulating spacers such as (BaO)Hg(BaO) is an order of magnitude smaller, but seems to have the same $\mathbf{k}_\parallel$-dependence. These hopping integrals are relevant for c-axis transport and their squares enter the Inter-layer Pair-Tunnelling model [19]. Then we show that the inclusion of $t'$ is crucial for understanding ARPES [20] for the anti-ferromagnetic insulator $Sr_2CuO_2Cl_2$. Finally, we estimate the value of the inter-plane exchange constant $J_\perp$ for a non-doped bi-layer in mean-field theory using different single-particle Hamiltonians, the LDA for $YBa_2Cu_3O_6$, the eight- and four-band Hamiltonians, as well as an analytical evaluation for the latter. We conclude that $J_\perp \sim -20$ meV.

## 2   LDA band structures of $YBa_2Cu_3O_7$ and $YBa_2Cu_4O_8$.

The full LDA band structure for $YBa_2Cu_3O_7$ [3] between the $\mathbf{k}_\parallel$ points $(0,0)$, $\left(\frac{\pi}{a}, 0\right)$, $\left(\frac{\pi}{a}, \frac{\pi}{b}\right)$, $\left(0, \frac{\pi}{b}\right)$, $(0,0)$, and $\left(\frac{\pi}{a}, \frac{\pi}{b}\right)$ is shown in Fig. 1 for $k_z=0$ (ΓXSYΓS), and in Fig. 2 for $k_z = \frac{\pi}{c}$ (ZURTZR). Also shown are the cross-sections of the Fermi surface with the $k_z=0$ and $\frac{\pi}{c}$ planes. Each of the small subfigures gives the projection onto a particular O $p$ or Cu $d$ orbital: The radius of the circle around the $\epsilon_n(\mathbf{k}_\parallel)$-point is proportional to the weight of that orbital in $\psi_n(\mathbf{k}_\parallel)$. Note that Cu and O orbitals have different normalizations in the figures so that weights cannot be compared *between* copper and oxygen. O3 are the plane-oxygens running along $b$, which is the direction of the chain. O2 are plane-oxygens along $a$, Cu2 are plane coppers, O4 apical- and O1 chain-oxygens, and Cu1 are chain-coppers.

The FS has four sheets denoted **a, b, c,** and **s.** The essential wave functions for the **a** and **b** sheets are respectively the **a**ntibonding (odd, -) and **b**onding (even, +) linear combinations for the bilayer of the *plane pdσ* anti-bonding orbitals ($O2_x$–$Cu2_{x^2-y^2}$–$O3_y$). Essential for the **c** sheet is the *chain pdσ* anti-bonding orbital made from $O1_y$ and the $pd\sigma$ anti-bonding dumbbell orbital ($O4_z$–$Cu1_{z^2-y^2}$–$O4_z$). Note that the orbitals denoted $z^2$ in the figures are $d_{3z^2-1}$ and that $d_{z^2-y^2} = \left(\sqrt{3}d_{3z^2-1} + d_{x^2-y^2}\right)/2$. The orbitals essential for the **s** sheet, the small stick pocket along SR, are two out of the three *chain pdπ* orbitals, which are the two $pd\pi$ anti-bonding dumbbell orbitals (O4$_x$–Cu1$_{xz}$–O4$_x$ and O4$_y$–Cu1$_{yz}$–O4$_y$) plus the anti-bonding linear combination of O1$_z$ with the $pd\pi$ bonding dumbbell orbital O4$_y$–Cu1$_{yz}$–O4$_y$. The anti-bonding linear combination of the two latter orbitals, which both involve O4$_y$, is the essential wave function for the stick.

If the one-electron wave functions stay coherent from chain to chain, as it is assumed in the LDA for a perfect crystal, then for $k_z=0$ there is neither hybridization between the **a** and **c** sheets, nor between the **b** and **s** sheets. Conversely, for $k_z=\frac{\pi}{c}$ there is neither hybridization between the **b** and **c** sheets nor between the **a** and **s** sheets. For a general $k_z$, all chain-plane hybridizations are allowed and give rise to a $k_z$-dispersion of several hundred meV's.

The Fermi level is approximately 20 meV above the odd plane-band saddle-points, which are bifurcated away from X and Y to respectively X$\pm\frac{1}{4}\Gamma$X and Y$\pm\frac{1}{3}\Gamma$Y. Since $b/a \approx 1.015$ and since the ratio between the Cu2$_{x^2-y^2}$–O3$_y$ and Cu2$_{x^2-y^2}$–O2$_x$ hopping integrals scales roughly as $(b/a)^{-4}$, we expect the saddle-point near X to lie more than 100 meV higher than the one near Y. However, the calculated energy difference is only 10 meV. The reason is that the odd plane band is repelled along $\Gamma$Y by the below-lying chain $pd\pi$ stick band (O4$_y$ and O1$_z$). Furthermore, the hybridization with the chain $pd\sigma$ band for $k_z \neq 0$ makes the plane-band saddle-points three-dimensional: The saddle-point related to $\left(\text{X} \pm \frac{1}{4}\Gamma\text{X}, -15 \text{ meV}\right)$ is (U, +80 meV) and the one related to $\left(\text{Y} \pm \frac{1}{3}\Gamma\text{Y}, -25 \text{ meV}\right)$ is $\left(\text{T} \pm \frac{1}{4}\Gamma\text{T}, -300 \text{ meV}\right)$. From the figures we see that not only the odd, but also the even plane band has bifurcated saddle-points, but these are about 0.5 eV below $\epsilon_F$.

Whether the above-mentioned $k_z$-dispersion persists in real YBa$_2$Cu$_3$O$_7$ is unclear: In the *double*-chain compound YBa$_2$Cu$_4$O$_8$, the LDA dispersion is an order of magnitude smaller than in YBa$_2$Cu$_3$O$_7$ because single-electron hopping through a double chain is frustrated by the fact that O1 in one chain plays the role of O4 for the other, and yet, the physical properties of the two compounds are quite similar.

In YBa$_2$Cu$_4$O$_8$ the LDA saddle-points of the odd plane band are (X$\pm\frac{1}{5}\Gamma$X, −20 meV), (U, 0 meV), (Y$\pm\frac{1}{4}\Gamma$Y, −160 meV) and (T, −180 meV). The difference between the energies near X and Y is thus as expected when the chain-plane hybridization is small and $b/a \approx 1.013$. The even-odd splitting of the plane bands caused by hopping across the double-layer is ∼0.5 eV near X or Y, exactly as in YBa$_2$Cu$_3$O$_7$.

## 3   8-band Hamiltonian.

The chain degrees of freedom have low energy, but they are not generic because not all HTCS's have chains. In order to derive a generic low-energy few-band Hamiltonian for the *planes*, we have therefore downfolded the LDA multi-band Hamiltonian for YBa$_2$Cu$_3$O$_7$ and $k_z=0$ to a two-center, orthogonal, tight-binding Hamiltonian $H_8^-\left(\mathbf{k}_\parallel\right)$ with eight *odd* bilayer orbitals interacting through nearest-neighbor hopping. Similarly, we have downfolded the LDA Hamiltonian for $k_z=\frac{\pi}{2}$ to an eight-band Hamiltonian $H_8^+\left(\mathbf{k}_\parallel\right)$ for the *even* bilayer orbitals [18]. Over a range of $\pm 1$ eV, $H_8^-$ and $H_8^+$ accurately describe the odd and even plane bands, respectively, as they are seen in Figs. 1 and 2. The weak hybridization with the chain $pd\pi$ bands has not been removed from, but is folded into these 8-band Hamiltonians.

The downfolding consisted of first integrating out all degrees of freedom with very high energies (e.g. Ba *spdf*, Y *spdf*, Cu *p*, O *sd*). This was done using the LMTO downfolding procedure (see also Sect. 4). Thereafter, all remaining orbitals other than the eight odd (even) *plane* oxygen and copper orbitals specified in the Introduction were deleted from the downfolded LMTO basis, and the odd (even) plane bands were recalculated. These eight-LMTO bands accurately reproduced the corresponding full LDA plane bands over a $\pm 2$ eV range. Finally, the non-orthogonality and long range of the eight odd (even) orbitals were transformed away by fitting the bands in the $\pm 1$ eV range to an orthogonal, two-center, nearest-neighbor, tight-binding Hamiltonian which then, is the one desired.

The reason for keeping in the Hamiltonian particularly those eight orbitals, is that these orbitals are the ones which after orthogonalization describe the LDA plane-bands accurately over a $\pm 1$ eV range with nearest-neighbor hoppings only, that is, with the minimal number of parameters. The 8-band Hamiltonian is thus the one which is "chemically" meaningful and sufficiently simple that further degrees of freedom may be integrated out analytically, as we shall demonstrate in the following section.

By transforming from orbitals even and odd with respect to the yttrium mid-plane of the bilayer to orbitals centered on a specific plane, one obtains a *single-plane* Hamiltonian, $H_8\left(\mathbf{k}_\parallel\right)$, plus integrals for hopping *between* the two planes of the bilayer.

Fig. 3 now *synthesizes* the eight plane-bands and gives the most significant parameters of $H_8$ (tetragonal averages are for instance taken). Ref. [18] provides the accurate values of *all* the matrix elements, as well as the detailed expressions, and the comparison with the full LDA bands. The inter-plane hopping will be discussed separately in Sect. 8.

Shown in the bottom left-hand panel of Fig. 3 are the four $\sigma$-orbitals looked upon from above the plane, $|y\rangle \equiv O3_y$, $|d\rangle \equiv Cu2_{x^2-y^2}$, $|x\rangle \equiv O2_x$, and $|s\rangle \equiv Cu2_s$. Below, we see, from the edge of the plane, the $|d\rangle$ orbital and two of the $\pi$-orbitals, $|z\rangle \equiv O2_z$ and $|xz\rangle \equiv Cu2_{xz}$. The orbital energies are given at the relevant points of the band structure shown in the left-hand panel, in eV and with respect to the energy of the $Cu2_{x^2-y^2}$ orbital. Hence, $\epsilon_x = \epsilon_y \equiv \epsilon_p$ is 0.9 eV *below* $\epsilon_d$, and $\epsilon_s$ is 6.5 eV above. Moreover, $\epsilon_{O2z} = \epsilon_{O3z} \equiv \epsilon_z$ is 0.4 eV above $\epsilon_d$, and $\epsilon_{xz} = \epsilon_{yz}$ is 1 eV below. If we now, as shown in the figures at the bottom, include the $pd\sigma$ hopping integrals $t_{xd} = t_{yd} \equiv t_{pd} = 1.6$ eV and the $pd\pi$ hopping integrals $t_{z,xz} = t_{z,yz} = 0.7$ eV, we obtain the band structure shown in the panel above: The $\sigma$-orbitals give rise to a bonding, a non-bonding, and an anti-bonding $O2_x$–$Cu_{x^2-y^2}$–$O3_y$ band plus a $Cu_s$ level (full lines), and the $\pi$-orbitals give rise to two decupled pairs of bonding anti-bonding bands which disperse in either the $x$ or $y$ direction (stippled lines). The anti-bonding $pd\sigma$ band, which will develop into the conduction band, has saddle-points at X (and Y) which are well above the top of the $\pi$-bands, and which are *isotropic* in the sense, that the absolute values of the band masses in the $x$ and $y$ directions are equal. This means that the FS at half-filling is a square with corners at X and Y.

In the middle panel, the $\sigma$ and $\pi$ bands remain decoupled, but we have introduced the $Cu_s$–$O2_x$ and $Cu_s$–$O3_y$ hoppings ($t_{sp} = 2.3$ eV), as well as the tiny $O3_z$–$O2_z$ hopping ($t_{zz} = 0.06$ eV). The latter is the only one reaching beyond nearest neighbors and it merely lifts a degeneracy of the anti-bonding $\pi$-bands at $\Gamma$. The *strong* coupling of the *remote* $Cu_s$ orbital to the $pd\sigma$ orbitals has the pronounced effect of depressing the conduction band near the saddle-points at X and Y, and thereby increasing the mass towards $\Gamma$ and decreasing it towards S $(\pi, \pi)$. The FS passing through X and Y will therefore correspond to finite hole-doping and will bulge towards $\Gamma$ (see the left-hand panel of Fig. 7). That the $Cu_s$ orbital, which has the azimuthal quantum number $m_z = 0$, may mix with the conduction band at X (and Y), but not at $\frac{1}{2}\Gamma S\left(\frac{\pi}{2}, \frac{\pi}{2}\right)$, can be seen from the symmetries of the corresponding two anti-bonding $pd\sigma$ wave functions shown schematically in Fig. 4. That there is, in fact, substantial by-mixing of $Cu_s$ character in the full LDA wave function at X, but none at $\frac{1}{2}\Gamma S$, is proved by the $|\psi(\mathbf{r})|^2$-contours shown in respectively the left and right-hand parts of Fig. 5. The by-mixing of $Cu_{3z^2-1}$ character (which also has $m_z = 0$-symmetry) to the so-called $Cu_s$ orbital, is seen to be small. The net downwards shift of the saddle-point energy is the result of down-pushing by pure $Cu_s$ and weak up-pushing by pure $Cu_{3z^2-1}$, whose energy is below the saddle-point. In Sect. 6 we shall give analytical descriptions of the conduction band in this $dp\sigma$-$sp\sigma$ approximation.

As a result of the hybridization with the $Cu_s$ orbital, the saddle-points of the anti-bonding $\sigma$-band straddle off the top of the appropriate $\pi$-band, so that even a weak dimple or buckle of the planes will introduce noticeable hybridization between the $\sigma$ and $\pi$ bands. This is seen in the right-hand panel where we have turned on the weak $Cu_{x^2-y^2}$–$O2_z$ and $Cu_{x^2-y^2}$–$O3_z$ hoppings ($t_{zd} = 0.24$ eV $\propto \sin\delta$). These couplings become allowed when there is a finite angle $\delta$ ($\approx 7^o$ in $YBa_2Cu_3O_7$ and $YBa_2Cu_4O_8$) between the Cu-O bond and the plane of the 2D-translations. Since the anti-bonding $pd\sigma$ and $Cu_{xz}$–$O2_z$ $pd\pi$ orbitals can only hybridize between $\Gamma$ and X, but not at X, the $\sigma$-$\pi$ hybridization makes the saddle-point bifurcate away from X.

In conclusion, the chemical eight-band Hamiltonian $H_8\left(\mathbf{k}_\parallel\right)$, whose eigenvalues give the one-

electron bands for a single plane, has the $\sigma$-block:

| $\langle\sigma|H|\sigma\rangle$ | $|\text{Cu }d\rangle$ | $|\text{Cu }s\rangle$ | $|\text{O2 }x\rangle$ | $|\text{O3 }y\rangle$ |
|---|---|---|---|---|
| $\langle\text{Cu }d|$ | $\epsilon_d$ | $0$ | $2t_{pd}\sin\frac{a}{2}k_x$ | $-2t_{pd}\sin\frac{a}{2}k_y$ |
| $\langle\text{Cu }s|$ | $0$ | $\epsilon_s$ | $2t_{sp}\sin\frac{a}{2}k_x$ | $2t_{sp}\sin\frac{a}{2}k_y$ |
| $\langle\text{O2 }x|$ | $2t_{pd}\sin\frac{a}{2}k_x$ | $2t_{sp}\sin\frac{a}{2}k_x$ | $\epsilon_p$ | $0$ |
| $\langle\text{O3 }y|$ | $-2t_{pd}\sin\frac{a}{2}k_y$ | $2t_{sp}\sin\frac{a}{2}k_y$ | $0$ | $\epsilon_p$ |

, (1)

the $\pi$-block:

| $\langle\pi|H|\pi\rangle$ | $|\text{O2 }z\rangle$ | $|\text{O3 }z\rangle$ | $|\text{Cu2 }xz\rangle$ | $|\text{Cu2 }yz\rangle$ |
|---|---|---|---|---|
| $\langle\text{O2 }z|$ | $\epsilon_z$ | $-4t_{zz}\cos\frac{a}{2}k_x\cos\frac{a}{2}k_y$ | $2t_{z,xz}\sin\frac{a}{2}k_x$ | $0$ |
| $\langle\text{O3 }z|$ | $-4t_{zz}\cos\frac{a}{2}k_x\cos\frac{a}{2}k_y$ | $\epsilon_z$ | $0$ | $2t_{z,xz}\sin\frac{a}{2}k_y$ |
| $\langle\text{Cu2 }xz|$ | $2t_{z,xz}\sin\frac{a}{2}k_x$ | $0$ | $\epsilon_{xz}$ | $0$ |
| $\langle\text{Cu2 }yz|$ | $0$ | $2t_{z,xz}\sin\frac{a}{2}k_y$ | $0$ | $\epsilon_{xz}$ |

,

and the block mixing the $\sigma$ and $\pi$ orbitals for dimpled planes:

| $\langle\sigma|H|\pi\rangle$ | $|\text{O2 }z\rangle$ | $|\text{O3 }z\rangle$ | $|\text{Cu2 }xz\rangle$ | $|\text{Cu2 }yz\rangle$ |
|---|---|---|---|---|
| $\langle\text{Cu }d|$ | $2t_{zd}\cos\frac{a}{2}k_x$ | $-2t_{zd}\cos\frac{a}{2}k_y$ | $0$ | $0$ |
| $\langle\text{Cu }s|$ | $0$ | $0$ | $0$ | $0$ |
| $\langle\text{O2 }x|$ | $0$ | $0$ | $0$ | $0$ |
| $\langle\text{O3 }y|$ | $0$ | $0$ | $0$ | $0$ |

## 4 Integrating out high-energy degrees of freedom.

In deriving the 8-band Hamiltonian from the LDA we have numerically integrated out many high-energy degrees of freedom. In the following, we shall integrate out further high-energy degrees of freedom starting from $H_8$ or $\langle\sigma|H|\sigma\rangle$. We now explain the Löwdin procedure:

Partitioning an orthonormal basis into $|i\rangle$ and $|j\rangle$, and down-folding the $H^{jj}$-block yields:

$$H_{ii'}(\epsilon) = H_{ii'} - \sum_{jj'} H_{ij}\left[\left(H^{jj} - \epsilon\right)^{-1}\right]_{jj'} H_{j'i'}. \qquad (2)$$

This is exact in the sense that $\det[H - \epsilon]$ and $\det[H^{ii}(\epsilon) - \epsilon]$ have identical zeroes, namely the eigenvalues of $H$. If we now Taylor-expand the down-folded Hamiltonian about $\epsilon_F$: $H^{ii}(\epsilon) = H^{ii}(\epsilon_F) + (\epsilon - \epsilon_F)\dot{H}^{ii}(\epsilon_F) + ...$, we see that the energy dependence of the down-folded Hamiltonian to linear order arises from non-orthogonality of the modified $i$-basis, the overlap matrix being $1 - \dot{H}^{ii}(\epsilon_F)$. The higher-order energy dependence of $H^{ii}(\epsilon)$ originates from energy dependence of the basis. The latter, and sometimes even the former energy dependencies can be neglected if the degrees of freedom to be integrated out have high energy, that is, if the energy range of interest lies far away from the eigenvalues of $H^{jj}$. An *orthonormal* low-energy Hamiltonian $\mathcal{H}$ may be obtained by orthogonalization of the modified $i$-basis, whose energy dependence has been neglected, e.g.

$$\mathcal{H} = \epsilon_F + \left[1 - \dot{H}^{ii}(\epsilon_F)\right]^{-\frac{1}{2}} \left[H^{ii}(\epsilon_F) - \epsilon_F\right] \left[1 - \dot{H}^{ii}(\epsilon_F)\right]^{-\frac{1}{2}}. \qquad (3)$$

## 5 3-band Hamiltonian.

A popular tight-binding Hamiltonian for flat planes is the 3-band Emery model. This is like $\langle\sigma|H|\sigma\rangle$ in (1), but without the Cu$_s$ orbital, with a renormalized oxygen energy $(\epsilon_p \to \epsilon'_p)$, and with 2nd-nearest-neighbor O2$_x$–O3$_y$ hopping.

Projecting the $Cu_s$ orbital out of $\langle\sigma|H|\sigma\rangle$ by means of (2) reveals that the values of $\epsilon_d$ and $t_{pd}$ are unchanged, but that 2nd-nearest-neighbor $O2_x$–$O3_y$ as well as 3rd-nearest-neighbor $O2_x$–$O2_x$ and $O3_y$–$O3_y$ hoppings, all of size $t_{pp} = t_{sp}^2/(\epsilon_s - \epsilon) \sim 1.1$ eV, must be added. Moreover, $\epsilon_p$ must be renormalized to $\epsilon'_p = \epsilon_p - 2t_{pp} \sim \epsilon_d - 3.0$ eV which is, in fact, the standard value. All of this comes about, because the $Cu_s$ character (see Fig. 5) is now built into the tails of the neighboring $O2_x$ and $O3_y$ orbitals.

According to the LDA, the Emery model should thus be modified to

$$\begin{array}{c|ccc}
H_3 & |Cu\,d\rangle & |O2\,x\rangle & |O3\,y\rangle \\
\langle Cu\,d| & \epsilon_d & 2t_{pd}\sin\frac{a}{2}k_x & -2t_{pd}\sin\frac{a}{2}k_y \\
\langle O2\,x| & 2t_{pd}\sin\frac{a}{2}k_x & \epsilon'_p + 2t_{pp}\cos ak_x & -4t_{pp}\sin\frac{a}{2}k_x\sin\frac{a}{2}k_y \\
\langle O3\,y| & -2t_{pd}\sin\frac{a}{2}k_y & -4t_{pp}\sin\frac{a}{2}k_x\sin\frac{a}{2}k_y & \epsilon'_p + 2t_{pp}\cos ak_y
\end{array} \quad (4)$$

Here we have neglected the non-orthogonality and, for the numerical values, we have taken the expansion energy, $\epsilon_F = \epsilon(X) \approx \epsilon_d + 1.53$ eV. As long as $\epsilon - \epsilon_F \ll \epsilon_s - \epsilon_F \sim 5$ eV, this is a good approximation. Orthogonalization by means of (3) would have introduced longer-range hoppings.

## 6 The $\sigma$-band.

In order to derive accurate and *transparent* expressions for the flat-plane $\sigma$-bands $\epsilon_n(\mathbf{k}_\parallel)$, or their constant-energy surfaces, we return to the 4-band Hamiltonian $\langle\sigma|H|\sigma\rangle$ in (1) and first Löwdin down-fold (2) the oxygen orbitals. The result is the following energy-dependent $sp\sigma$-$dp\sigma$ Hamiltonian:

$$\begin{array}{c|cc}
H_{sd}(\epsilon) & |Cu\,d,\epsilon\rangle & |Cu\,s,\epsilon\rangle \\
\langle Cu\,d,\epsilon| & \epsilon_d + (1-u)(2t_{pd})^2/(\epsilon - \epsilon_p) & -v(2t_{pd})(2t_{sp})/(\epsilon - \epsilon_p) \\
\langle Cu\,s,\epsilon| & -v(2t_{pd})(2t_{sp})/(\epsilon - \epsilon_p) & \epsilon_s + (1-u)(2t_{sp})^2/(\epsilon - \epsilon_p)
\end{array} \quad (5)$$

where we have defined the *new* $\mathbf{k}_\parallel$-*coordinates*:

$$u \equiv \frac{1}{2}(\cos ak_y + \cos ak_x) \quad \text{and} \quad v \equiv \frac{1}{2}(\cos ak_y - \cos ak_x), \quad (6)$$

both limited to the range from $-1$ to $+1$ and sketched in the left-hand side of Fig. 6. The origin of the $uv$-system is at $\frac{1}{2}\Gamma S$ and corresponding values of $(k_x, k_y)$ and $[u, v]$ are: $\frac{1}{2}\Gamma S\left(\frac{\pi}{2a}, \frac{\pi}{2a}\right)[0, 0]$, $X\left(\frac{\pi}{a}, 0\right)[0, 1]$, $\Gamma(0, 0)[1, 0]$, and $S\left(\frac{\pi}{a}, \frac{\pi}{a}\right)[-1, 0]$. Readers used to think about pairing and gap symmetries, will recognize $u$ as $s$- and $v$ as $d$-wave symmetry. The relation to the $\mathbf{k}_\parallel$-coordinates $x$ and $y$ used in Ref. [18] is: $1 - u = x + y$ and $v = x - y$.

The $sd$-Hamiltonian (5) highlights the $\mathbf{k}_\parallel$-dependence of the LDA $\sigma$-bands for a flat $CuO_2$ plane as follows: Without $s$-mixing, the bands depend only on $u$ (the saddle-points are isotropic, the FS for half filling is the square: $u = 0$, etc.) and the mixing element is proportional to $v$.

Further Löwdin down-folding, this time of the $sp\sigma$-block, yields the following energy-dependent 1-band Hamiltonian:

$$H(\epsilon) = \epsilon_d + \frac{(2t_{pd})^2}{\epsilon - \epsilon_p}\left(1 - u - \frac{v^2}{1 + s(\epsilon) - u}\right), \quad (7)$$

which is still equivalent with $\langle\sigma|H|\sigma\rangle$. When there is no mixing with Cu $s$, that is when $t_{sp} = 0$, $s(\epsilon) \to \infty$ so that the term proportional to $v^2$ vanishes. The quadratic $sp\sigma$-scattering function

$$s(\epsilon) \equiv (\epsilon_s - \epsilon)(\epsilon - \epsilon_p)/(2t_{sp})^2 \quad (8)$$

is fairly constant in the neighborhood of $\epsilon = (\epsilon_s + \epsilon_p)/2 \approx 2.8$ eV, where it takes its maximum value $[(\epsilon_s - \epsilon_p)/4t_{sp}]^2 \approx 0.65$. For instance is $s(1.5\text{ eV}) \approx 0.57$.

The most elegant way of expressing the eigenvalues of $\langle \sigma | H | \sigma \rangle$ is as the roots of the secular equation,

$$[H(\epsilon) - \epsilon] \frac{\epsilon - \epsilon_p}{(2t_{pd})^2} \equiv 1 - u - \frac{v^2}{1 + s(\epsilon) - u} - d(\epsilon) = 0, \tag{9}$$

because from its definition,

$$d(\epsilon) = (\epsilon - \epsilon_d)(\epsilon - \epsilon_p)/(2t_{pd})^2 \tag{10}$$

is a quadratic function of energy which depends only on the $dp\sigma$ scattering. For (9) to vanish, we see that $d(\epsilon)$ must increase from 0 to 2 in the region of the conduction band because $s(\epsilon)$ is positive and because the bottom of the band is at $\Gamma$ and the top is at S. For a given energy, $\epsilon$, we calculate the values of $s(\epsilon)$ and $d(\epsilon)$ from (8) and (10), whereafter the requirement that (9) be zero gives the exact constant-energy surface. For a given $\mathbf{k}_\parallel$, on the other hand, we can not give an exact, explicit expression for the four eigenvalues, $\epsilon_n(\mathbf{k}_\parallel)$, but near the middle of the conduction band we may exploit the near-constancy of $s(\epsilon)$ and are then led by (10) and (9) to solve the following equation:

$$\epsilon(\mathbf{k}) = \frac{\epsilon_p + \epsilon_d}{2} + \sqrt{\left(\frac{\epsilon_p - \epsilon_d}{2}\right)^2 + (2t_{pd})^2 \left(1 - u - \frac{v^2}{1 + s(\epsilon) - u}\right)} \tag{11}$$

iteratively [i.e. $\epsilon(\mathbf{k}) \to \epsilon$]. For ease of notation, we have dropped the subscript $\parallel$ on $\mathbf{k}_\parallel$, and we shall do so from now on.

With the parameter-values from Fig. 3, we easily obtain: $\epsilon(X) = \epsilon(u=0, v=1) \approx 1.53$ eV and $\epsilon\left(\frac{1}{2}\Gamma S\right) = \epsilon(u=0, v=0) \approx 2.78$ eV. The $Cu_s$ level thus pushes the saddle-point down by 1.25 eV. The energy of the bottom of the conduction band, $\epsilon(\Gamma) = \epsilon(u=1, v=0) = \max(\epsilon_p, \epsilon_d) = 0$, and of its top, $\epsilon(S) = \epsilon(u=-1, v=0) \approx 4.10$ eV, are independent of $s$.

The $Cu_d$, $Cu_s$, and $O_p$ (O2+O3) characters of the conduction-band wave function are most easily obtained as the respective derivatives $\partial \epsilon(\mathbf{k})/\partial \epsilon_d$, $\partial \epsilon(\mathbf{k})/\partial \epsilon_p$, and $\partial \epsilon(\mathbf{k})/\partial \epsilon_s$. This follows from 1st-order perturbation theory. By differentiation of (9) and making use of

$$\dot{d}(\epsilon) = 2\frac{\epsilon - (\epsilon_p + \epsilon_d)/2}{(2t_{pd})^2} \quad \text{and} \quad \dot{s}(\epsilon) = 2\frac{(\epsilon_p + \epsilon_s)/2 - \epsilon}{(2t_{sp})^2}. \tag{12}$$

we obtain the simple results:

$$|c_d(\epsilon, \mathbf{k})|^2 = \frac{1}{2} \frac{\epsilon - \epsilon_p}{\epsilon - (\epsilon_p + \epsilon_d)/2} \left[1 - \frac{\dot{s}(\epsilon)}{\dot{d}(\epsilon)} \frac{v^2}{[1 + s(\epsilon) - u]^2}\right]^{-1} \tag{13}$$

and

$$\left|\frac{c_s(\epsilon, \mathbf{k})}{c_d(\epsilon, \mathbf{k})}\right|^2 = \left(\frac{t_{pd}}{t_{sp}}\right)^2 \frac{v^2}{[1 + s(\epsilon) - u]^2} \tag{14}$$

and $|c_p(\epsilon, \mathbf{k})|^2 = 1 - |c_s(\epsilon, \mathbf{k})|^2 - |c_d(\epsilon, \mathbf{k})|^2$. The $Cu_d$ character is seen to be $\sim \frac{1}{2}[1 - ..]^{-1}$ when the energy is far away from $\epsilon_p \sim \epsilon_d$. Moreover, the *variation* of Cu $d$ character along a constant-energy surface, e.g. the FS, becomes negligible when

$$1 \gg \left|\frac{\dot{s}(\epsilon)}{\dot{d}(\epsilon)}\right| = \left(\frac{t_{pd}}{t_{sp}}\right)^2 \left|\frac{\epsilon - (\epsilon_p + \epsilon_s)/2}{\epsilon - (\epsilon_p + \epsilon_d)/2}\right| \approx 0.48 \left|\frac{\epsilon - 2.8 \text{ eV}}{\epsilon + 0.45 \text{ eV}}\right|, \tag{15}$$

and vanishes for the energy midway between the $O_p$ and $Cu_s$ energies. With our parameter-values and for $\epsilon$=1.53 eV, $\dot{s}/\dot{d} \approx 0.31$, whereas for $\epsilon$=2.78, $\dot{s}/\dot{d} = 0$. In the following, we shall often neglect $\dot{s}$, and when this can be done, the $Cu_d$ character is independent of $\mathbf{k}$ and slightly larger than 0.5. The $Cu_s$ character is proportional to $v^2$ and, with our values of the hopping integrals and our $s$-value, $|c_s(\epsilon, \mathbf{k})|^2 \sim 0.12v^2(1 - 0.6u)^{-2}$, that is, at X there is 61% $Cu_d$, 12% $Cu_s$, and 27% $O2_x$ character.

# 7  1-band Hamiltonians.

The 1-band Hamiltonian for an orbital which is orthogonal to itself when displaced by a lattice translation, i.e. for a Wannier function, has the **k**-dependence of a Fourier series:

$$\begin{aligned}\mathcal{H}\left(\mathbf{k}_{\parallel}\right) &= \langle\epsilon\rangle - 2t\left(\cos ak_x + \cos ak_y\right) + 4t'\cos ak_x\cos ak_y - 2t''\left(\cos 2ak_x + \cos 2ak_y\right) \\ &\quad + 4t^{(3)}\left(\cos ak_x\cos 2ak_y + \cos ak_y\cos 2ak_x\right) + 4t^{(4)}\cos 2ak_x\cos 2ak_y \\ &\quad - 2t^{(5)}\left(\cos 3ak_x + \cos 3ak_y\right) + \ldots - 2t^{(9)}\left(\cos 4ak_x + \cos 4ak_y\right) + \ldots\end{aligned} \qquad (16)$$

with $t^{(n)}$ being the hopping integral between $(n+1)$st nearest neighbors and $\langle\epsilon\rangle$ being the average energy of the band.

The conduction band shown in the right-hand panel of Fig. 3, which is given by the 8-band Hamiltonian for a plane dimpled by 7°, is well separated from all other bands, and this band is sufficiently smooth that its Fourier series converges at a reasonable pace. The results obtained numerically by diagonalizing $H_8^-$ and $H_8^+$ (with the parameters given in Ref. [18]) and subsequent Fourier transformation of the odd and even conduction bands are given in the first two rows of the following table:

| | $\langle\epsilon\rangle$ | $t$ | $t'$ | $t''$ | $t^{(3)}$ | $t^{(4)}$ | $t^{(5)}$ | $t^{(9)}$ |
|---|---|---|---|---|---|---|---|---|
| $H_8^-(\mathbf{k})/$meV | 140 | 349 | 96 | 62 | 18 | 1 | 10 | 1 |
| $H_8^+(\mathbf{k})/$meV | $-140$ | 422 | 113 | 110 | 20 | 5 | 32 | 11 |
| $\varepsilon(\mathbf{k})\times\dot{d}$ | | $\frac{1}{4}\left(1+\frac{1}{8}r\right)$ | $\frac{1}{4}r(1+\frac{1}{2}r^2)$ | $\frac{1}{8}r(1+\frac{1}{4}r^2)$ | $\frac{1}{16}r^2$ | $\frac{1}{16}r^3$ | $\frac{1}{16}r^2$ | $\frac{1}{32}r^3$ |

The corresponding even and odd Wannier orbitals must have fairly long range. The zero of energy in the table is at the average of the even and odd bands. The splitting between these bands will be discussed in Sect. 8.

With decreasing dimpling, the convergence of the Fourier series (16) of the entire conduction band deteriorates even further. Reasons are the near-crossing of the $\sigma$ and $\pi$-bands, and the near-cusp of the $\sigma$-band at $\Gamma$. The latter can be seen in the middle panel of Fig. 3 and is caused by the near degeneracy of $\epsilon_d$ and $\epsilon_p$. For the pure $\sigma$-band, a more useful low-energy Hamiltonian is therefore obtained by Fourier transformation, not of the *entire* anti-bonding band, but merely of the part near $\epsilon_F$. We shall now give a simple, analytical derivation of such a low-energy $\sigma$-band Hamiltonian. The result is given in the last row of the table with $r\equiv 1/[2(1+s)]$. This Hamiltonian, $\varepsilon(\mathbf{k})$, reproduces the $\sigma$ conduction band to linear order in $\epsilon - \epsilon_F$ and Eq. (22) below gives the second-order correction.

Let us start with the remark, that Eq. (11) with $s(\epsilon)\equiv s(\epsilon_F)$ is in fact a highly accurate 1-band Hamiltonian for the $\sigma$ conduction band; it even describes the cusp behavior near $\Gamma$ where $[u,v]\sim[1,0]$. However, (11) is not an *orthonormal* 1-band Hamiltonian (16), because such a Hamiltonian must depend *analytically* on $u$ and $v$. To obtain a 1-band Hamiltonian analytical in $u$ and $v$, we first expand $d(\epsilon)$ and $s(\epsilon)$ around $\epsilon_F$. If we only go to first order, we can solve Eq. (9) and obtain for the 1st-order estimate:

$$\varepsilon(\mathbf{k}) = \epsilon_F + \left(1 - d - u - \frac{v^2}{1+s-u}\right)\left[\dot{d} - \dot{s}\frac{v^2}{[1+s-u]^2}\right]^{-1} \qquad (17)$$

$$\approx \epsilon_F + \dot{d}^{-1}\left[1 - d - u - 2rv^2\left\{1 + 2ru + (2ru)^2 + \ldots\right\}\right]. \qquad (18)$$

Note the difference between $\varepsilon$ and $\epsilon$. Here, $d\equiv d(\epsilon_F)$, $\dot{d}\equiv\dot{d}(\epsilon_F)$, $s\equiv s(\epsilon_F)$, $r\equiv 1/[2(1+s)]$ and $\dot{s}\equiv\dot{s}(\epsilon_F)$. As an alternative to (12) we may use:

$$\dot{d}^{-1} = \frac{t_{pd}}{\sqrt{d + [(\epsilon_p - \epsilon_d)/4t_{pd}]^2}} \sim \frac{t_{pd}}{\sqrt{d}}. \qquad (19)$$

In (18) we have *neglected* $\dot{s} \equiv \dot{s}(\epsilon_F)$ and *expanded* $1/(1+s-u)$ for small $u$. Substitution of the cosines (6) for $u$ and $v$ and comparing terms with (16) yields the hopping integrals listed in the last row of the table given above. Here the zero of energy is taken at the average band energy

$$\langle \epsilon \rangle = \epsilon_F + \dot{d}^{-1}\left[1 - d - \frac{1}{2}r(1 + \frac{1}{2}r^2)\right] \tag{20}$$

and

$$r \equiv \frac{1/2}{1+s} \approx 0.32, \quad \text{so that} \quad \frac{t'}{t} \approx r \quad \text{and} \quad \frac{t''}{t'} \approx \frac{1}{2}. \tag{21}$$

For a flat plane and *without* $s$-hybridization ($r=0$), the 1-band 1st-order Hamiltonian $\varepsilon(\mathbf{k})$ only has nearest neighbor interactions. For realistic $s$-values, hopping to second *and third* neighbors must be included, but farther hoppings can be neglected. In previous many-body calculations, $t'$ was often neglected, but when it was included, $t' \sim 0.3t$ was in fact a standard value. $t''$ has sofar been neglected.

Had we kept $\dot{s}/\dot{d}$ to lowest order, we should have added to the coefficient of $2rv^2$ in (18) the term $2r(1-d-u)\dot{s}/\dot{d}$. Moreover, $r$ in the table and in (20) should be substituted by $r' \equiv r\left[1 - 2r(1-d)\dot{s}/\dot{d}\right]$.

As an example, let us take $\epsilon_F$ at the saddle-point, $[u,v] = [0,1]$, then $\epsilon_F \approx 1.53$ eV, as found from (11), $r \approx 0.32$ from (21), $d = 1 - 2r \approx 0.363$ from (9), $\dot{d}^{-1} \approx 2.64$ eV, from (19), $2r(1-d)\dot{s}/\dot{d} \approx 0.13$ and $r' = 0.28$. Hence, from the table, $t \approx 0.69$ eV, $t'/t \approx 0.28$, and $t''/t' \approx 0.49$. Using the first-order expression (18) to calculate the energy at $\frac{1}{2}\Gamma S$, which we know from (11) should be 1.25 eV above the saddle-point, we find: $\varepsilon\left(\frac{1}{2}\Gamma S\right) - \epsilon_F = (1-d)/\dot{d} = 2r'\dot{d}^{-1} \approx 1.46$ eV. This discrepancy is due to the lack of higher-order terms in (11).

The second-order corrections to the 1st-order Hamiltonian $\varepsilon(\mathbf{k})$ may be obtained by expanding $d(\epsilon)$ to second order and using (9) and (18) to obtain:

$$\epsilon(\mathbf{k}) + \frac{1}{2}\frac{[\epsilon(\mathbf{k}) - \epsilon_F]^2}{\epsilon_F - (\epsilon_p + \epsilon_d)/2} = \varepsilon(\mathbf{k}).$$

This is the exact expression for $\epsilon(\mathbf{k})$ (still assuming $s$ constant) because $d(\epsilon)$ is a quadratic function and $\epsilon_F - (\epsilon_p + \epsilon_d)/2 = d/\dot{d}$. For the purpose of deriving an analytical 1-band 2nd-order Hamiltonian, we do not want to solve this equation because the result is the exact Eq. (11), written as:

$$\frac{\epsilon(\mathbf{k}) - \epsilon_F}{\epsilon_F - (\epsilon_p + \epsilon_d)/2} = -1 + \sqrt{1 + 2\frac{\varepsilon(\mathbf{k}) - \epsilon_F}{\epsilon_F - (\epsilon_p + \epsilon_d)/2}}.$$

To obtain an analytical Hamiltonian, we must expand as follows:

$$\epsilon(\mathbf{k}) - \epsilon_F = [\varepsilon(\mathbf{k}) - \epsilon_F]\left[1 - \frac{1}{2}\frac{\varepsilon(\mathbf{k}) - \epsilon_F}{\epsilon_F - (\epsilon_p + \epsilon_d)/2} + ...\right] \tag{22}$$

Since $\epsilon_F - (\epsilon_p + \epsilon_d)/2 \sim 2$ eV, this expansion hardly converges for $\varepsilon\left(\frac{1}{2}\Gamma S\right) - \epsilon_F = 1.46$ eV, but only for energies much closer to $\epsilon_F$. This is the cusp-problem, showing up again. Our 1-band 1st-order Hamiltonian including 3rd-nearest-neighbor interactions is therefore appropriate for energies one order of magnitude closer to $\epsilon_F$ than $(\epsilon_p + \epsilon_d)/2$, i.e. below 200 meV. To develop an expression more explicit than (22) for a 1-band 2nd-order Hamiltonian which treats higher energies better than $\varepsilon(\mathbf{k})$, seems pointless considering the complication of squaring the Fourier series for $\varepsilon(\mathbf{k}) - \epsilon_F$.

We should also comment on the most obvious way to derive the $\sigma$ conduction-band Hamiltonian, namely linearization and orthonormalization (3) of $H(\epsilon)$ in (7). Using (9) and (18) it becomes obvious that

$$\mathcal{H}(\mathbf{k}) - \epsilon_F = \frac{H(\epsilon_F) - \epsilon_F}{1 - \dot{H}(\epsilon_F)} = \frac{\varepsilon(\mathbf{k}) - \epsilon_F}{1 + (\varepsilon(\mathbf{k}) - \epsilon_F)/(\epsilon_F - \epsilon_p)} \equiv [\varepsilon(\mathbf{k}) - \epsilon_F]\left[1 - \frac{\varepsilon(\mathbf{k}) - \epsilon_F}{\epsilon_F - \epsilon_p} + ...\right].$$

Comparison with the correct expansion (22) shows that $\mathcal{H}(\mathbf{k})$ is merely correct to first order, like $\varepsilon(\mathbf{k})$, but it has longer range. In conclusion, $\varepsilon(\mathbf{k})$ seems to be the 1st-order 1-band Hamiltonian with the simplest $\mathbf{k}$-dependence.

Finally we remark that for the 8-band model, too, a 1-band 1st-order Hamiltonian may be derived from Ref. [18] where it is shown that the constant energy contour for an arbitrary energy is a polynomial of second order in $x = \frac{1}{2}(1 - \cos ak_x)$ and in $y = \frac{1}{2}(1 - \cos ak_y)$. By expansion of the coefficients [A($\epsilon$) through I($\epsilon$)] to linear order in $\epsilon - \epsilon_F$ and solving, we find $\epsilon - \epsilon_F$ to be given by the *ratio* of two polynomials. Expansion of the denominator gives the orthonormal 1-band Hamiltonian.

# 8  Inter-plane hopping, $t_\perp \left( \mathbf{k}_\| \right)$.

From the parameters of $H_8^-$ and $H_8^+$ given in [18] for YBa$_2$Cu$_3$O$_7$ we derive the values of the integrals

$$t_{ii}^\perp = \frac{1}{2}\left( \epsilon_{ii}^- - \epsilon_{ii}^+ \right) \quad \text{and} \quad t_{ij}^\perp = \pm \frac{1}{2}\left( t_{ij}^+ - t_{ij}^- \right)$$

for hopping from orbital $i$ in the lower plane to orbital $j$ in the upper plane of the bilayer. We chose the signs of the hopping integrals in such a way that the integrals are positive if the two orbitals have pure cubic-harmonic character and are strongly localized. In the expression for $t_{ij}^\perp$, the upper sign is therefore chosen for $\sigma$-$\sigma$ and $\pi$-$\pi$ inter-layer hopping integrals, and the lower sign is chosen for $\sigma$-$\pi$ hopping. The (tetragonally averaged and rounded) values of the inter-plane hopping integrals are:

| $t_{ss}^\perp$ | $t_{sp}^\perp$ | $t_{pp}^\perp$ | $t_{dd}^\perp$ | $t_{zz}^\perp$ | $t_{zx,zx}^\perp = t_{zy,zy}^\perp$ | $t_{z,zx}^\perp = t_{z,zy}^\perp$ | |
|---|---|---|---|---|---|---|---|
| 0.75 | 0.27 | 0.12 | 0.05 | 0.35 | $-0.30$ | $-0.15$ | eV |

The negative signs presumably indicate that the corresponding hoppings mostly proceed via yttrium and barium.

For the conduction band, the most important inter-plane hoppings are the vertical $t_{ss}^\perp$ from Cu$_s$ to Cu$_s$ and $t_{sp}^\perp$ from Cu$_s$ in one plane to the nearest O2$_x$ or O3$_y$ in the other plane. This is so, even though the Cu$_s$ character in the conduction band is relatively small. This statement may be verified from the left-hand side of Fig. 7, where we show

$$t_\perp(\mathbf{k}) \equiv \frac{1}{2}\left[ \epsilon^-(\mathbf{k}) - \epsilon^+(\mathbf{k}) \right], \tag{23}$$

with the odd and even conduction bands calculated by numerical diagonalization of respectively $H_4^-$ and $H_4^+$ in (1), including all parameters as given in Ref. [18]. The even-odd splitting is seen almost to vanish along $\Gamma$M [M$\equiv$S$=\left(\frac{\pi}{a},\frac{\pi}{a}\right)$] and to reach its maximum of 0.6 eV at X. This qualitatively follows the $v^2/(1-2ru)^2$ dependence of the Cu$_s$ character (14). Below, we shall work this out in detail. Also shown in the figure (heavy lines) are the constant-energy contour (FS) of the odd band passing through the saddle-point and the two neighboring ones corresponding to energies $\pm 100$ meV from the saddle-point.

The right-hand side of Fig. 7 shows the same even-odd splitting and the odd-band constant-energy contours at, and $\pm 100$ meV from, the saddle-point, but now calculated from $H_8^-$ and $H_8^+$. The 8-band Hamiltonian gives a more appropriate description of the even-odd splitting in the dimpled CuO$_2$ bi-layer (See Figs. 1 and 2 for YBa$_2$Cu$_3$O$_7$). The effects of the relatively strong hopping between $\pi$-orbitals on different planes are noticeable, but nevertheless, the even-odd splitting still attains its minimum along $\Gamma$M and its maximum of 0.6 eV at the saddle-point. That the maximum of $t_\perp(\mathbf{k})$ moves along with the saddle-point as it bifurcates due to dimpling of the plane, is caused by vertical O$_z$-O$_z$ hopping.

We now give an analytical description of the essentials of $t_\perp(\mathbf{k})$ for a flat plane. Hence, we consider only the $\sigma$-band and neglect all inter-plane hoppings other than $t_{ss}^\perp$ and $t_{sp}^\perp$. In this case, the Hamiltonians are $H^P(\epsilon)$ in (7), which depend on the mid-plane parity ($P$) only through $s^P(\epsilon)$. We therefore immediately realize that the inter-plane hopping is *proportional to* $v^2$. From (18):

$$t_\perp(\mathbf{k}) = \frac{t_{pd}^2}{\epsilon_F - (\epsilon_p + \epsilon_d)/2} \left( \frac{v^2}{1 + s^+ - u} - \frac{v^2}{1 + s^- - u} \right) \approx t_\perp(X) \left( \frac{v}{1 - 2ru} \right)^2 \quad (24)$$

where the maximum is at X [$u=0, v=1$] and takes the value

$$t_\perp(X) = 4r^2 \left( s^- - s^+ \right) \frac{t_{pd}^2}{\epsilon_F - (\epsilon_p + \epsilon_d)/2} \approx 8r^2 s \left( \frac{t_{ss}^\perp}{\epsilon_s - \epsilon_F} + 2 \frac{t_{sp}^\perp}{t_{sp}} \right) \frac{t_{pd}^2}{\epsilon_F - (\epsilon_p + \epsilon_d)/2} \quad (25)$$

As usual, $s \equiv s(\epsilon_F)$, $r \equiv [2(1+s)]^{-1} \sim t'/t$, and $\dot{s}$ has been neglected. Taking again $\epsilon_F$ at the X-point of the un-split band, we get: $t_\perp(X) \approx 8 \cdot 0.32^2 \cdot 0.57 \left( \frac{0.75}{6.5 - 1.53} + 2\frac{0.27}{2.3} \right) \frac{1.6^2}{1.53 + 0.45}$eV $\approx 0.23$ eV. Inclusion of $\dot{s}$ reduces this value by 13% to 0.20 eV. Had we instead taken $\epsilon_F = \epsilon\left(\frac{\pi}{2a}, \frac{\pi}{2a}\right) = 2.78$ eV, we would have had $\dot{s} = 0$, $s = 0.65$, and $r = 0.30$, so that $t_\perp(X) \approx 0.17$ eV. Both estimates are still the main part of the 0.29 eV obtained when including also those inter-plane hoppings which do not involve $Cu_s$. Note that the inter-plane $Cu_s$-$O_p$ hopping contributes as much to $t_\perp(\mathbf{k})$ as $Cu_s$-$Cu_s$. This is because $O_p$ is a major part of the conduction-band wave function. (One might then wonder about the importance of $t_{pp}^\perp$ and $t_{dd}^\perp$, but their effects tend to cancel). The fact that $t_\perp(\mathbf{k})$ in the left-hand side of Fig. 7 is not symmetric around the line XX is mainly due to the factor $(1-2ru)^{-2}$ in Eq. (24).

Inspection of LDA calculations for other multi-layer systems reveals that the $t_\perp(\mathbf{k})$-behavior described above is generic, as we would expect on the grounds that the $Cu_s$ (and $Cu_{3z^2-1}$) orbital is always present and the inter-layer distance is fairly constant. What might change, are of course $\epsilon_s$ and $t_{sp}$ because they depends on the balance between the $Cu_s$ and $Cu_{3z^2-1}$ characters which could be offset if the distance to apical oxygen became unusually short, that is, more than 0.1 Å shorter than the 2.30 Å found in $YBa_2Cu_3O_7$. The bandstructures calculated by Novikov et al. [22] for bi-layer Hg-1212 and tri-layer Hg-1223, which have flat $CuO_2$ planes separated by Ca, reveal that $t_\perp$ has $v^2$-dispersion (24) and that $t_\perp(X) \sim 0.25$ eV, like for undimpled $YBa_2Cu_3O_7$. Exactly the same is found from Mattheiss and Hamann's calculation [23] for flat bi-layer $Bi_2Sr_2CaCu_2O_8$ (provided that one looks at the bands where they are purely $CuO_2$-like). Finally, all calculations for the idealized infinite-layer material $CaCuO_2$ with flat planes and without apical oxygen exhibit $k_z$-dispersion like $2t_\perp(\mathbf{k}_\parallel) \cos ck_z$, exactly as expected from (24). The maximum bandwidth is 1 eV so that, here again, $t_\perp(X) \sim 0.25$ eV.

A related, but less generic problem is the inter-plane hopping through insulating spacer layers, such as (BaO)Hg(BaO) in the Hg compounds. Here, inspection of both existing calculations [22, 24] for the single-layer material reveals that the $k_z$-dispersion is like $2t_\perp(\mathbf{k}_\parallel) \cos ck_z$, once more, and that $t_\perp(X) \approx 30$ meV in both calculations, that is, *an order of magnitude smaller* than for hopping between multi-layers. Considering the fact, that this "backwards hopping" proceeds mainly through the apical-oxygen $z$-orbital, $v^2$-behavior is to be expected, because the $O4_z$-orbital has $m_z=0$ and therefore only couples to $Cu_s$.

$t_\perp(\mathbf{k})$ is the important quantity in the inter-layer pair tunnelling mechanism of P.W. Anderson et al. [19] for boosting $T_c$, and $v^2$-dependence was in fact assumed by Chakravarty inspired by LDA bands. According to the inter-layer pair tunnelling theory, correlation effects block the hopping of single electrons between the planes, but in the superconducting state a Cooper pair may tunnel due to the Josephson coupling:

$$H_J \approx -\sum_\mathbf{k} T_J(\mathbf{k}) \left( c_{\mathbf{k}\uparrow}^{(1)\dagger} c_{-\mathbf{k}\downarrow}^{(1)\dagger} c_{-\mathbf{k}\downarrow}^{(2)} c_{\mathbf{k}\uparrow}^{(2)} + h.c. \right) \quad \text{where} \quad T_J(\mathbf{k}) \approx t_\perp(\mathbf{k})^2 / t.$$

BCS mean-field theory then yields the following crude estimate of the anisotropy of the superconducting gap and $T_c$:

$$\Delta(\mathbf{k}_F) \sim \Delta_0(\mathbf{k}_F) + T_J(\mathbf{k}_F)/2 \quad \text{and} \quad T_c \sim \max T_J(\mathbf{k}_F)/4$$

With our parameter values, this means that the gap for a bi-layer should be something like $\Delta_0(\mathbf{k}_F) + 50$meV$\cdot v^4/(1-2ru)^4$, traced along the FS (see Fig. 7), and that $T_c \sim 300$ K.

# 9 Band dispersion in an anti-ferromagnetic insulator.

It has been a challenge to understand how the valence band, or rather, its spectral density, evolves as a function of doping as the material changes from an anti-ferromagnetic insulator (AFI) to a metal, becoming superconducting at low temperature. ARPES [20] for the un-dopable AFI $Sr_2CuO_2Cl_2$ with a single flat $CuO_2$ plane, gave a valence-band which dispersed little from $\Gamma$ $(0,0)$ to X $(\pi,0)$ and which rose to a nearly isotropic maximum at $\frac{1}{2}\Gamma S$ $\left(\frac{\pi}{2},\frac{\pi}{2}\right)$ (see Fig. 6). Except for the band-width, this did not agree with many-body calculations for the $t-J$ [9] and $t-U$ [25] Hubbard models, which predicted the rise to be from $\Gamma$ to X, and the maximum to be flat along the AF zone-boundary $(\pi,0)$-$\left(\frac{\pi}{2},\frac{\pi}{2}\right)$.

In the right-hand side of Fig. 6 we show the calculated band structure of an AFI in the simplest possible approximation where an external staggered field of size $\pm\frac{m}{2}U$ was applied to the $Cu_{x^2-y^2}$ orbital. The results for $Um = 2, 4,$ and 6 eV are shown. In order to reproduce the experimental gap of 1.8 eV, one would choose $Um \sim 5$ eV. The single-particle Hamiltonian was taken to be the 4-band Hamiltonian $\langle\sigma|H|\sigma\rangle$. One observes that the calculated dispersion has exactly the *same* characteristics as the experimental one, except that the band-width is 4 times too large, as might have been expected. We made the points [26] that the discrepancy between the experiment and the many-body calculations is caused by the use of an oversimplified single-particle Hamiltonian, namely the 1-band nearest-neighbor Hamiltonian (the $t$-model), and that inclusion of $t'$ (and $t''$) should fix the problem. This has turned out to be true [27]. Let us now explain the mean-field results in some detail.

For $Um = 0$ (not shown), the bands are simply the anti-bonding $\sigma$-band folded into the AF Brillouin zone. This folding is around the $v$-axis, so that $u$ and $-u$ become equivalent, and the two bands are degenerate along the $v$-axis. The dispersion along the $v$-axis is the 1.25 eV that we calculated in connection with Eq. (11) and which, on the $d$-scale, is from 1 to $1-2r$. For the $t$-model this dispersion does not exist. The relevant feature seen in Fig. 6 is that this dispersion along the AF zone boundary hardly changes when the staggered field is turned on. All that happens is that a gap of about half the size of the field ($\sim 50\%$ $Cu_d$ character) opens up while the dispersion essentially remains.

The mathematics of this is really simple: The staggered field operates on $Cu_d$ and couples $\epsilon(\mathbf{k})$ to $\epsilon(\mathbf{k}+\mathbf{q})$ with $\mathbf{q} \equiv \left(\frac{\pi}{a},\frac{\pi}{a}\right)$, that is, $\epsilon(u,v)$ is coupled to $\epsilon(-u,v)$. The energy dependent 1-band $\sigma$-Hamiltonian (7) therefore becomes:

$$\begin{array}{c|cc}
H(\epsilon) & |\mathbf{k}\rangle & |\mathbf{k}+\mathbf{q}\rangle \\
\hline
\langle\mathbf{k}| & \epsilon_d + \frac{(2t_{pd})^2}{\epsilon-\epsilon_p}\left(1 - u - \frac{v^2}{1-u+s(\epsilon)}\right) & \frac{m}{2}U \\
\langle\mathbf{k}+\mathbf{q}| & \frac{m}{2}U & \epsilon_d + \frac{(2t_{pd})^2}{\epsilon-\epsilon_p}\left(1 + u - \frac{v^2}{1+u+s(\epsilon)}\right)
\end{array} \qquad (26)$$

Along the AF zone boundary, $u=0$, we obtain:

$$\epsilon = \epsilon_d + \frac{(2t_{pd})^2}{\epsilon-\epsilon_p}\left(1 - \frac{v^2}{1+s(\epsilon)}\right) \pm \frac{m}{2}U$$

This demonstrates that the dispersion along the $v$-axis is independent of the staggered field, and that the splitting is independent of $v$. Hence, the $-2rv^2$ dispersion remains whatever it was for $U=0$. These statements are of course only true if we neglect the energy-dependences of $\epsilon-\epsilon_p$ and $s(\epsilon)$; in reality the valence band tends to get compressed at energies closer to $\epsilon_p$ and to gain $O_p$ character. Solving for $\epsilon$ yields (11) with $u=0$ and $\epsilon_d$ substituted by $\epsilon_d \pm \frac{m}{2}U$.

For an energy range which is narrow in comparison with the distances from $\epsilon_p \sim \epsilon_d$ and $\epsilon_s$, we may like in (18) expand the diagonal of the secular matrix $[H(\epsilon) - \epsilon](\epsilon-\epsilon_p)/(2t_{pd})^2$ to linear order around an energy at the center of interest. For consistency, we shall denote this energy $\epsilon_F$, although it will usually not be the Fermi energy. In order that the resulting 2×2 Hamiltonian yields eigenvalues correct to linear order, the off-diagonal matrix element needs only be correct at $\epsilon_F$. For simplicity, we shall neglect $\dot{s}$, which is a reasonable approximation for the un-split and for the *upper* Hubbard band because its center is at ~4.1 eV when $Um \sim 5$ eV. The resulting 1st-order Hamiltonian is simply:

$$
\begin{array}{c|cc}
\mathcal{H} & |\mathbf{k}\rangle & |\mathbf{k+q}\rangle \\
\hline
\langle\mathbf{k}| & \varepsilon(\mathbf{k}) & \frac{m}{2}U|c_d|^2 \\
\langle\mathbf{k+q}| & \frac{m}{2}U|c_d|^2 & \varepsilon(\mathbf{k+q})
\end{array},
\tag{27}
$$

where $\varepsilon(\mathbf{k})$ was given by (18) and $|c_d|^2 = (\epsilon_F - \epsilon_p)(2t_{pd})^{-2}\dot{d}^{-1} = \frac{\epsilon_F - \epsilon_p}{2\epsilon_F - \epsilon_p - \epsilon_d} \sim \frac{1}{2}$ is the $\mathrm{Cu}_d$ character (13) at $\epsilon_F$. As previously discussed, $|c_d|^2$ does not depend on $\mathbf{k}$ when $\dot{s} = 0$. Had we gone to higher order in the $\epsilon - \epsilon_F$ expansion, the Hamiltonian would still have been (27), but with $\varepsilon(\mathbf{k})$ and $|c_d|^2$ substituted by respectively $\epsilon(\mathbf{k})$ and $|c_d(\mathbf{k})|^2$. This form holds generally, e.g. also if it were derived from the 8-band model and on-site Coulomb interactions were included on other orbitals, when we define:

$$
U|c(\mathbf{k})|^2 \equiv \langle\psi(\mathbf{k})|U|\psi(\mathbf{k+q})\rangle = \sum c_i^*(\mathbf{k})U_i c_i(\mathbf{k+q}).
\tag{28}
$$

From (27), the upper and lower bands are then:

$$
E(\mathbf{k}) = \frac{\epsilon(\mathbf{k}) + \epsilon(\mathbf{k+q})}{2} \pm \sqrt{\left(\frac{\epsilon(\mathbf{k}) - \epsilon(\mathbf{k+q})}{2}\right)^2 + \left(\frac{m}{2}U|c(\mathbf{k})|^2\right)^2}.
\tag{29}
$$

If $\epsilon(\mathbf{k})$ and $|c(\mathbf{k})|^2$ are low-energy expansions, the result of (29) only applies close to $\epsilon_F$. Hence, we must use different $\epsilon_F$'s for the upper and lower bands, as is also natural, because their characters are very different when $Um \sim 5$ eV: The valence band is $\mathrm{O}_p$-like and the upper Hubbard band is $\mathrm{Cu}_d$-like. Remember also, that $H(\epsilon)$ in (26) for an anti-ferromagnet describes 8 rather than 2 bands, and that (29) can describe any of them through choice of $\epsilon_F$ and the appropriate sign.

## 10 Inter-plane exchange across a bi-layer, $J_\perp$.

For anti-ferromagnetic bi-layer materials, such as $\mathrm{YBa_2Cu_3O_6}$, the spin-order across the bi-layer is anti-ferromagnetic and the inter-layer exchange-coupling constant $J_\perp$ is an important parameter in many theories of high-temperature superconductivity and for understanding the origin of the so-called spin gap [28, 29]. From neutron scattering experiments on $\mathrm{YBa_2Cu_3O_{6+x}}$ it has been found [30] that, upon hole doping far into the metallic regime, the anti-ferromagnetic spin-correlations between the planes are more persistent that those in the planes. The value of $J_\perp$ is under dispute, but it seems clear that in $\mathrm{YBa_2Cu_3O_6}$ the absolute value must exceed 7 meV because the optical spin-wave branch, which should be at energy $2\sqrt{J_\parallel J_\perp}$, could not be detected with neutron scattering up to 60 meV [31], and the intra-plane exchange constant is experimentally known to be –120 meV. From mid-infrared measurements of the spin-wave spectrum in $\mathrm{YBa_2Cu_3O_6}$ [32] it was recently concluded that $J_\perp \sim -65$ meV. SEDOR experiments on the $\mathrm{YBa_2Cu_3O_7}$-$\mathrm{YBa_2Cu_4O_8}$ *metallic* compound [33], on the other hand, indicates that $J_\perp/J_\parallel$ increases strongly with decreasing temperature and reaches a maximum value of 0.3 just above $T_c$. This presumably puts an upper bound of 50 meV on $J_\perp$. We shall now estimate the value of $J_\perp$ through several implementations of simple mean-field theory. In all cases we assume the value $Um = 5$ eV for the self-consistent field applied to the Cu $d$ orbitals.

We consider the insulating phase, apply the staggered field $\pm 2.5$ eV inside each plane, and calculate the tiny difference, $\mathcal{E}(\mathrm{F}) - \mathcal{E}(\mathrm{AF})$, in the sum of the band-structure energies according to whether the orientation of the staggered field is ferro- (F) or anti-ferromagnetic (AF) between the to planes of the bi-layer. This energy difference per Cu spin is $J_\perp/4$. Before we present an analytical calculation based on the results of the two previous sections, we list results of various numerical calculations:

| Hamiltonian: | $H_8$ | LDA | $H_8$ | $H_4$ | $H_4\, t_{ss}^\perp, t_{sp}^\perp$ | $H_4\, t_{ss}^\perp, t_{sp}^\perp$ |
|---|---|---|---|---|---|---|
| Procedure: | $4\langle t_\perp^2/U\rangle$ | F,AF | F,AF | $4\langle t_\perp^2/U\rangle$ | $4\langle t_\perp^2/U\rangle$ | F,AF |
| $U$: | $\mathrm{Cu}_{x^2-y^2,xz,yz}$ | $\mathrm{Cu}_{x^2-y^2}$ | $\mathrm{Cu}_{x^2-y^2}$ | $\mathrm{Cu}_{x^2-y^2}$ | $\mathrm{Cu}_{x^2-y^2}$ | $\mathrm{Cu}_{x^2-y^2}$ |
| $J_\perp/\mathrm{meV}$: | $-25$ | $-13$ | $-13$ | $-17$ | $-8$ | $-6$ |

These result are seen to scatter between −6 and −25 meV. Our largest calculation (LDA) used the charge-selfconsistent LDA potential for non-magnetic YBa$_2$Cu$_3$O$_6$ and applied the staggered field to the Cu$_{x^2-y^2}$ orbital in a standard multi-orbital band-structure calculation using the orthonormal LMTO representation, diagonalization of the Hamiltonian, and summation over the occupied bands in the AF Brillouin zone to obtain $\mathcal{E}(\mathrm{F})$ and $\mathcal{E}(\mathrm{AF})$. Surprisingly, the same result ($J_\perp = -13$ meV) was obtained with the same brute-force procedure (F,AF), but using $H_8^+$ and $H_8^-$ for the non-magnetic part of the Hamiltonian. In the latter calculation we took advantage of the fact that the magnetic perturbation for inter-plane F/AF-order is even/odd so that it can not/only mix $H^+$ and $H^-$. In this way, the matrices to be diagonalized were only 16×16 for $H_8$. If we apply the same procedure, but start from the simplified $t_{ss}^\perp, t_{sp}^\perp$-version of $H_4$ which includes inter-plane hopping via Cu $s$ orbitals only, we find a $J_\perp$ which is only −6 meV. The three remaining numerical calculations listed in the table employed a perturbative procedure labelled $4\langle t_\perp^2/U\rangle$, which we shall now explain.

We use the 1-band expression (29) and calculate the change in the total energy as the *negative* of the change of the total energy of merely the *upper* Hubbard band. This neglects the change of the energy of the highest, Cu$_s$-like band. Expansion of (29) to lowest order in $[\epsilon(\mathbf{k}) - \epsilon(\mathbf{k}+\mathbf{q})]/Um|c(\mathbf{k})|^2$ (the result (30) will hold beyond this order) yields for the upper band:

$$E(\mathbf{k}) \approx \frac{\epsilon(\mathbf{k}) + \epsilon(\mathbf{k}+\mathbf{q})}{2} + \frac{m}{2}U|c(\mathbf{k})|^2 + \frac{1}{mU|c(\mathbf{k})|^2}\left(\frac{\epsilon(\mathbf{k}) - \epsilon(\mathbf{k}+\mathbf{q})}{2}\right)^2$$

For a bi-layer, inter-plane F-order couples $\epsilon^\pm(\mathbf{k})$ to $\epsilon^\pm(\mathbf{k}+\mathbf{q})$, while inter-plane AF-order couples $\epsilon^\pm(\mathbf{k})$ to $\epsilon^\mp(\mathbf{k}+\mathbf{q})$. The even and odd upper bands for F-order are therefore:

$$E^\pm(\mathbf{k}) \approx \frac{\epsilon^\pm(\mathbf{k}) + \epsilon^\pm(\mathbf{k}+\mathbf{q})}{2} + \frac{m}{2}U|c(\mathbf{k})|^2 + \frac{1}{mU|c(\mathbf{k})|^2}\left(\frac{\epsilon^\pm(\mathbf{k}) - \epsilon^\pm(\mathbf{k}+\mathbf{q})}{2}\right)^2,$$

while the two upper bands for AF-order are

$$E^{1,2}(\mathbf{k}) \approx \frac{\epsilon^\pm(\mathbf{k}) + \epsilon^\mp(\mathbf{k}+\mathbf{q})}{2} + \frac{m}{2}U|c(\mathbf{k})|^2 + \frac{1}{mU|c(\mathbf{k})|^2}\left(\frac{\epsilon^\pm(\mathbf{k}) - \epsilon^\mp(\mathbf{k}+\mathbf{q})}{2}\right)^2.$$

The difference in total energy per Cu spin thus works out as:

$$-\frac{1}{4}J_\perp \equiv \mathcal{E}(\mathrm{F}) - \mathcal{E}(\mathrm{AF}) = \frac{1}{2}\left\langle E^1(\mathbf{k}) + E^2(\mathbf{k}) - E^+(\mathbf{k}) - E^-(\mathbf{k})\right\rangle = \left\langle\frac{t_\perp(\mathbf{k})\, t_\perp(\mathbf{k}+\mathbf{q})}{mU|c(\mathbf{k})|^2}\right\rangle, \tag{30}$$

where $\langle\rangle$ denotes the average over the Brillouin zone. As usual (23), $t_\perp(\mathbf{k})$ is half the splitting between the odd and even bands for $U=0$. Expression (30) in real space is, in fact, the familiar one for Anderson super-exchange.

The results labelled $4\langle t_\perp^2/U\rangle$ were now obtained by diagonalizing the appropriate non-magnetic 4×4 or 8×8 Hamiltonians $H^+(\mathbf{k})$ and $H^-(\mathbf{k})$ numerically, extracting the odd-even splitting to form $t_\perp(\mathbf{k})$, the eigenvectors to form $mU|c(\mathbf{k})|^2$, and summing over the Brillouin-zone (30). The result obtained in this way for the simplified 4-band Hamiltonian ($J_\perp = -8$ meV) is in accord with, but slightly larger than that obtained with the direct procedure for the same Hamiltonian. The result obtained for the full 4-band Hamiltonian ($J_\perp = -17$ meV) is twice as large. Finally, since the conduction band of the 8-band model has Cu$_{xz}$ and Cu$_{yz}$ components besides Cu$_{x^2-y^2}$, we did a calculation in which $|c(\mathbf{k})|^2$ denoted the sum of the characters for all three Cu $d$-orbitals, rather than just that of $Cu_{x^2-y^2}$. The corresponding decrease of the denominator in (30) raised $J_\perp$ to –25 meV.

We thus conclude that, for the purpose of calculating $J_\perp$ in the mean-field approximation, $H_8$ simulates the full LDA Hamiltonian very well, the perturbation-expression (30) for $J_\perp$ is accurate, the assumption of inter-plane hopping via only $Cu_s$ under-estimates $J_\perp$ by a factor 2, by-mixing of Cu $d\pi$ character enhances $J_\perp$, and our *best estimate* of $J_\perp$ is ~–20 meV.

With expression (30) we may evaluate $J_\perp$ *analytically* for $H_4$, using (13) for the $Cu_{x^2-y^2}$ character and the 1st-order expressions (24) and (25) for $t_\perp(\mathbf{k})$ in the $t^\perp_{ss}, t^\perp_{sp}$-approximation. We obtain:

$$-J_\perp = 4\frac{t_\perp(X)^2}{mU|c_d|^2}\left(\frac{3}{8}\right)^2\left(1+\frac{2}{9}r^2+..\right) \approx 4\frac{(0.17\text{ eV})^2}{5\text{ eV} \cdot 0.57} \cdot 0.14 = 6\text{ meV} \tag{31}$$

where $t_\perp(X)$ was given in (25) and the last factors come from the Brillouin-zone average:

$$\left\langle\left(\frac{v^2}{1-(2ru)^2}\right)^2\right\rangle = \left\langle v^4\left[1+2(2ru)^2+..\right]\right\rangle = \left(\frac{3}{8}\right)^2\left(1+\frac{2}{9}r^2+..\right).$$

The numerical estimate for $-J_\perp$ was obtained by using the expansion energy $\epsilon_F = \epsilon\left(\frac{\pi}{2a}, \frac{\pi}{2a}\right) = 2.78$ eV, which is at the center of the U=0-band, and for which we can safely neglect $\dot{s}$. The –6 meV is in good agreement with the values obtained numerically for the same Hamiltonian. Since the $v^4$-dependence dominates the $ru$-dependence in the average over the Brillouin-zone, expression (31) for $J_\perp$ is fairly insensitive to details of the dispersion. If for $t_\perp(X)$ we merely insert the LDA value of 0.25 eV, expression (31) in fact yields: $J_\perp \approx -13$ meV, which is the LDA result.

Barriquand and Sawatzky [34] recently obtained $J_\perp \sim -56$ meV from a real-space evaluation of (30). One reason for the discrepancy with our value is, that these authors only considered the perpendicular hopping integral $t^\perp_{pp}$ and took it to be 2.5 times larger than our value (as given in Sect. 8). It seems to us, that had the perpendicular hopping been dominated by $t^\perp_{pp}$, then $t_\perp(\mathbf{k})$ would have been nearly independent of $\mathbf{k}$, in contradiction with the even-odd splittings found in all LDA calculations.

In a numerical (F,AF)-calculation, we can simulate hole-doping crudely by moving the Fermi level into the valence band. At the same time, the mean field should be reduced in such a way that it vanishes at a hole doping of about 0.2 per plane. Fig. 8 shows the result of such a calculation of $J_\perp$ for many different values of the field. The persistence of the inter-plane anti-ferromagnetic coupling with increasing doping is clearly seen, until a point where $J_\perp$ changes sign and the coupling becomes ferro-magnetic.

## 11 Acknowledgments

We are grateful to T.M. Rice for suggesting that we evaluate $J_\perp$ from our low-energy Hamiltonian. Discussions with Z.-X. Shen, J.C. Campuzano, W. Hanke, D.J. Scalapino, and P.W. Anderson are gratefully acknowledged.

**FIGURE CAPTIONS**

**Fig. 1.** LDA energy bands, orbital projected bands, and Fermi surface for $YBa_2Cu_3O_7$ and $k_z$=0 [3].

**Fig. 2.** Same as Fig. 1, but for $k_z = \frac{\pi}{c}$.

**Fig. 3.** Specification of the 8-band Hamiltonian and synthesis of its band structure. All energies are in eV. See text.

**Fig. 4.** $O2_x$-$Cu_{x^2-y^2}$-$O3_y$ conduction-band states at $\mathbf{k} = \left(\frac{\pi}{a}, 0\right)$ and $\left(\frac{\pi}{2a}, \frac{\pi}{2a}\right)$, schematic.

**Fig. 5.** LDA wave-functions $\left(\text{surfaces of constant } |\psi(\mathbf{k}, \mathbf{r})|^2\right)$ for the odd plane conduction band in $YBa_2Cu_3O_7$. *Left:* $\mathbf{k} = \left(\frac{\pi}{a}, 0, 0\right)$ X. *Right:* $\mathbf{k} = \left(\frac{\pi}{2a}, \frac{\pi}{2a}, 0\right) \frac{1}{2}\Gamma S$ [21].

**Fig. 6.** *Left:* Non-magnetic and anti-ferromagnetic Brillouin zones and relation between the $(k_x, k_y)$ and $[u, v]$ coordinate systems. *Right:* Energy bands from the 4-band Hamiltonian in the presence of anti-ferromagnetically staggered fields acting on the $\text{Cu}_{x^2-y^2}$ orbital and of strengths $\pm 1$, $\pm 2$, and $\pm 3$ eV.

**Fig. 7.** Contours of constant $t_\perp(\mathbf{k}) \equiv \frac{1}{2}[\epsilon^-(\mathbf{k}) - \epsilon^+(\mathbf{k})]$ calculated for the 4- and 8-band models. The heavy lines give the odd constant energy surfaces passing through the saddle-points, as well as the two neighboring contours corresponding to the energies $\pm 100$ meV from the saddle-point.

**Fig. 8.** $J_\perp$ as a function of hole doping calculated for many different values of the staggered field. The single-particle Hamiltonian was $H_4\, t_{ss}^\perp, t_{sp}^\perp$ and the procedure was (F,AF) as explained in the text.